\documentclass[twocolumn]{aastex7}

\usepackage{hyperref}
\usepackage{amsmath}
\usepackage{cleveref}
\usepackage{latexsym,amssymb}
\usepackage{graphicx}
\usepackage{array}
\usepackage{comment}
\usepackage{caption}
\usepackage{subcaption}
\usepackage{multirow}

\makeatletter
\usepackage{etoolbox}
\patchcmd\H@refstepcounter{\protected@edef}{\protected@xdef}{}{}
\makeatother

\newcommand{\GN}{\ensuremath{G_{\rm N}}}

\newcommand{\cnfw}{\ensuremath{c_{200}^{\rm{NFW}}}}

\newcommand{\Munit}{\ensuremath{10^{14}\, M_{\odot}}}
\newcommand{\Md}{M_{\Delta}}
\newcommand{\cd}{c_{\Delta}}
\newcommand{\Rd}{R_{\Delta}}

\newcommand{\Mf}{M_{200}}
\newcommand{\rf}{r_{200}}
\newcommand{\cf}{c_{200}}

\newcommand{\Mcau}{M_{200}^{\rm Cau}}
\newcommand{\Mnfw}{M_{200}^{\rm NFW}}
\newcommand{\Mcauf}{M_{\rm Cau}}
\newcommand{\Mnfwf}{M_{\rm NFW}}
\newcommand{\Ngal}{N_{\rm gal}}
\newcommand{\Nclus}{N_{\rm clu}}
\newcommand{\fofR}{f(\mathcal{R})}

\newcommand{\rp}{r_{\rm p}}
\newcommand{\phitwo}{\phi_{\rm \infty, 2}}
\newcommand{\betatwo}{\beta_{\rm 2}}

\newcommand{\ksM}{\text{km/s Mpc$^{-1} $}}
\newcommand{\ks}{\text{km s$^{-1}$}}
\newcommand{\Psz}{\ensuremath{P_{\rm SZ}}}

\newcommand{\Tx}{\ensuremath{T_{\rm X}}}
\newcommand{\fbr}{\mathcal{F}_{\beta}(r)}
\newcommand{\fb}{\mathcal{F}_{\beta}}
\newcommand{\gbr}{g_{\beta}(r)}
\newcommand{\fr}{\mathcal{F}(r)}
\newcommand{\Mpc}{\rm {Mpc}}

\newcommand{\diff}{\rm {d}}
 
\newcommand{\LCDM}{\text{$\Lambda$CDM} }
\newcommand{\fR}{|f_{\rm R0}| }

\graphicspath{{./}{Plots_Paper/}}

\shorttitle{Testing gravity with stacked galaxy clusters}
\shortauthors{M.A. Butt et al.}

\begin{document}

\nolinenumbers

\title{Outer regions of galaxy clusters as a new probe to test modifications to gravity}

\author[0009-0006-7255-0484]{Minahil Adil Butt}
\email[show]{mbutt@sissa.it}
\affiliation{SISSA, Via Bonomea 265, 34136 Trieste, Italy}
\affiliation{IFPU - Institute for fundamental physics of the Universe, Via Beirut 2, 34014 Trieste, Italy}
\affiliation{INFN-Sezione di Trieste, via Valerio 2, 34127 Trieste, Italy}
\affiliation{ICTP-The Abdus Salam International Centre for Theoretical Physics, Strada Costiera 11, 34151 Trieste, Italy  }

\author[0000-0002-9153-1258]{Sandeep Haridasu} 
\email[show]{sharidas@sissa.it}
\affiliation{SISSA, Via Bonomea 265, 34136 Trieste, Italy}
\affiliation{IFPU - Institute for fundamental physics of the Universe, Via Beirut 2, 34014 Trieste, Italy}
\affiliation{INFN-Sezione di Trieste, via Valerio 2, 34127 Trieste, Italy}

\author[0000-0002-4986-063X]{Antonaldo Diaferio}
\affiliation{Dipartimento di Fisica, Università di Torino, via P. Giuria 1, 10125, Torino, Italy}
\affiliation{Istituto Nazionale di Fisica Nucleare, Sezione di Torino, via P. Giuria 1, 10125, Torino, Italy}
\email[show]{antonaldo.diaferio@unito.it}

\author{Francesco Benetti}
\affiliation{SISSA, Via Bonomea 265, 34136 Trieste, Italy}\affiliation{IFPU - Institute for fundamental physics of the Universe, Via Beirut 2, 34014 Trieste, Italy}\affiliation{INFN-Sezione di Trieste, via Valerio 2, 34127 Trieste, Italy}
\email[]{fbenetti@sissa.it}

\author{Yacer Boumechta}
\affiliation{SISSA, Via Bonomea 265, 34136 Trieste, Italy}\affiliation{IFPU - Institute for fundamental physics of the Universe, Via Beirut 2, 34014 Trieste, Italy}\affiliation{INFN-Sezione di Trieste, via Valerio 2, 34127 Trieste, Italy}
\affiliation{ICTP-The Abdus Salam International Centre for Theoretical Physics, Strada Costiera 11, 34151 Trieste, Italy  }
\email[]{}

\author{Carlo Baccigalupi}
\affiliation{SISSA, Via Bonomea 265, 34136 Trieste, Italy}\affiliation{IFPU - Institute for fundamental physics of the Universe, Via Beirut 2, 34014 Trieste, Italy}\affiliation{INFN-Sezione di Trieste, via Valerio 2, 34127 Trieste,  Italy}
\email[show]{\\ bacci@sissa.it}

\author[0000-0002-4882-1735]{Andrea Lapi}
\affiliation{SISSA, Via Bonomea 265, 34136 Trieste, Italy}\affiliation{IFPU - Institute for fundamental physics of the Universe, Via Beirut 2, 34014 Trieste, Italy}\affiliation{INFN-Sezione di Trieste, via Valerio 2, 34127 Trieste,  Italy}\affiliation{IRA-INAF, Via Gobetti 101, 40129 Bologna, Italy}
\email[show]{lapi@sissa.it}

\begin{abstract}

We apply the caustic technique to samples of galaxy clusters stacked in redshift space to estimate the gravitational potential in the cluster's outer region and test modifications to the standard theory of gravity. We separate 122 galaxy clusters from the HeCS-SZ, HeCS-redMapper, and HeCS samples into four samples with increasing mass; we estimate four robust, highly constraining caustic profiles for these samples. The caustic masses of the four stacked clusters agree within $ 10\%$  with the corresponding median values of each cluster sample. By adopting the NFW density profile to model the gravitational potential, we recover the caustic profile $\mathcal{A}(r)$ up to radius $r_{\rm p}  \sim 4.0\, {\rm Mpc}$. This comparison is a first-order validation of the mass-concentration relation for galaxy clusters expected in the $\Lambda$CDM model. We thus impose this correlation as a prior in our analysis. Based on our stacked clusters, we estimate the value of the filling factor, which enters the caustic technique, $\fb = 0.59\pm 0.05$; we derive this value using real data alone and find it consistent with the value usually adopted in the literature. We then use the caustic profiles $\mathcal{A}(r)$ of the stacked clusters to constrain the chameleon gravity model. We find that the caustic profiles provide a stringent upper limit of $|f_{\rm R0}| \lesssim 4 \times 10^{-6}$ at 95\% C.L. limits in the $f(\mathcal{R})$ scenario. The formalism developed here shall be further refined to test modifications to gravity in the extended outer weak gravitational regions of galaxy clusters. 

\end{abstract}


%%%%%%%%%%%%%%%%%%%%%%%%%%%%%%%%%%%%%%%%%%
\section{Introduction}\label{sec:intro}  
%%%%%%%%%%%%%%%%%%%%%%%%%%%%%%%%%%%%%%%%%%

The accurate measurement of galaxy cluster masses plays a pivotal role in our pursuit of precision cosmological measurements \citep{Hallman:2005zs, Pierre_2011, Reed:2012ih, Crone_1997}. Several ongoing efforts are aimed at understanding and managing both statistical \citep{Cromer:2021iyp, Tchernin:2020vbt, Melin_2015, K_hlinger_2015} and systematic uncertainties \citep{K_hlinger_2015, Rasia:2006ra} in the techniques used for cluster mass inference through direct measurements of gravitational potential \citep{Tchernin:2020vbt, Tamura:2000bd, Liang:2000yx}. Accurately measuring cluster potentials typically requires highly constraining data, whether it involves the number of background galaxies for weak lensing shear profiles \citep{Chen_2020, Schneider:1999ch, King:2000wc, Umetsu:2015baa}, X-ray temperature or SZ-pressure  for gas mass profiles \citep{Tamura:2000bd, Ettori:2018tus, Ettori:2013tka, Ettori10, Maughan_2016, Perotto:2023fyb}, or spectroscopic redshifts of galaxies for phase-space analyses to obtain the caustic profile \citep{Diaferio_1997, Diaferio99,Maughan_2016, Geller:2012ds}. 

The caustic method has proven to be an extremely successful technique for estimating the total mass of galaxy clusters, completely independent of the dynamical state. It relies on the estimation of the escape velocities based on the projected galaxy phase space. The caustic surface, which traces the amplitude of the phase space of the galaxies, facilitates the reconstruction of the total enclosed mass of the cluster, which extends far beyond the virial radius \citep{Diaferio99, Serra_2010, Gifford13, Logan_2022}. In turn, it is this ability to assess the mass profile beyond the virial radius that provides us with a unique observable to test galaxy cluster dynamics, estimate the extended gravitational potential and subsequently perform tests of gravity. In this context, an earlier proof-of-concept analysis was performed in \cite{Butt:2024jes}, who assessed the  mass bias present between the hydrostatic technique, based on data from \citet{Ettori:2018tus, Eckert:2018mlz}, and the caustic technique, for two well-observed galaxy clusters, A2029 and A2142 \citep{Sohn_2017, Liu_2018}. 

The caustic technique is a robust method for estimating the masses of galaxy clusters, as it is independent of the dynamical state of the cluster \citep{Diaferio_1997, Maughan_2016}. This is particularly important because the dynamical state of the cluster can significantly affect the mass estimates obtained from hydrostatic methods \citep{Biffi:2016yto}. The caustic technique also does not require any assumption for the density profile of the cluster, once the clusters from by hierarchical clustering, making it a powerful tool to estimate galaxy cluster masses in an almost model-independent way. Indeed, utilizing the caustic method requires a calibration constant, the filling factor $\fbr$, which is usually estimated with N-body simulations \citep{Diaferio_2005, Gifford13}. In our previous work \citep{Butt:2024jes}, we implement the complete expression of $\fbr$ assuming the NFW density/potential profile, and obtain competitive constraints on the parameters of modified gravity models. Unlike in \citet{Butt:2024jes}, where we performed a joint analysis of the caustic and hydrostatic mass, here we aim to obtain a more robust caustic surface(s) by stacking galaxy clusters, allowing us to obtain better constraints of the gravitational potential, independent of other mass estimation techniques. 
 
In estimating the caustic surface accurately, the quality and quantity of data can negatively impact the mass estimates when considering large samples of faint or small clusters with very few well-observed galaxies \citep{Gifford13}. To address this issue, we use stacking to enhance the signal-to-noise ratio for an ensemble of galaxy clusters complied in the HeCS-omnibus sample \citep{Sohn_2020}, providing strongly constrained mass estimates. HeCS-omnibus combines and extends three surveys: HeCS-SZ \citep{Rines_2016}, HeCS-redmapper \citep{Rines_2018} and HeCS \citep{Rines_2013}. Stacking also standardizes the projected shapes of the clusters, thereby also minimizing biases that arise from fitting spherical profiles to non-spherical systems \citep{Gifford:2016plw}. In the present work, we start by building  stacked galaxy clusters for which the galaxy phase space has been well-observed \citep{Rines_2016, Rines_2018, Sohn_2020}. 
% Furthermore, , here we employ the caustic technique alone, circumventing the need for hydrostatic data. 

The caustic technique has previously been applied to stacked systems in observational studies: for example, \citet{Biviano_2003} stacked 43 clusters with galaxies up to $2r_{\mathrm{vir}}$ and employed both the caustic technique and a Jeans analysis \citep{Binney1987} to recover an average mass profile, finding good agreement between the two methods. Similarly, \citet{Rines_2003} created an ensemble cluster based on nine clusters from the CAIRNS survey \citep{Rines:2005ta}. Unlike other studies, each of the included systems was sampled sufficiently to obtain individual measurements of velocity dispersion and virial mass ($M_{200}$). They scaled their velocities by the velocity dispersion of each system before stacking, resulting in agreement within $1\sigma$ of the theoretical expectation of $M_{200} = 3\sigma^{2}r_{200}/\GN$. \citet{Gifford13} developed and tested a new stacking algorithm based on the caustic phase-space itself, which reduces the mass scatter in $<\mathrm{ln} (M_{\mathrm{caustic}}|M_{200})>$ for ensemble clusters from $70\%$ for individual clusters to less than $10\%$ for ensemble {clusters with only 15 galaxies per cluster and 100 clusters per ensemble. }

In summary, our focus in this work is to use the caustic method, independently of other observables, on stacked galaxy clusters to obtain NFW-fitted \citep{navarro97} cluster masses. We further use this method to obtain constraints on the modified gravity model, chameleon screening \citep{Khoury:2003aq, Terukina:2013eqa}. We also observe and present the effect of including a concentration-mass prior on our mass fits and the resulting $\fbr$ and $g_{\beta}(r)$ obtained from our fitted parameters.  

The organization of the paper is as follows: We present the compilation of the data in \Cref{sec:data} followed by the methodology in \Cref{sec:method}. Finally, we present our results, alongside a discussion in \Cref{sec:results} and conclude in \Cref{sec:conclusions}. We assume $H_0 = 70 \, \ksM$ and $\Omega_{\rm m} = 0.3$ when estimating the critical density $\rho_{\rm{crit}} = 8 \pi \GN / 3 H^2(z) $ within the $\LCDM$ model with $H^2(z) = \Omega_{\rm m} (1+z)^3 + 1- \Omega_{\rm m}$.

%%%%%%%%%%%%%%%%%%%%%%%%%%%%%%%%%%%%%%%%%%
\section{Caustic Technique} \label{sec:Caustic}
%%%%%%%%%%%%%%%%%%%%%%%%%%%%%%%%%%%%%%%%%%
In this section, we present a brief description of the caustic technique primarily used to estimate the mass of galaxy clusters. As introduced in \citet[see also \citealt{Diaferio99, Diaferio09, Serra_2010}]{Diaferio_1997}, the caustic technique is a model-independent (i.e, without an explicit assumption of a density profile) approach to estimate the mass of galaxy clusters, utilizing the escape velocity of its member galaxies inferred from the projected phase space. The key significance of the method is that the mass profile can be estimated up to large radii, a few times the virial radius ($\rf$) of the cluster, and is independent of the dynamical state of the cluster. The distribution of the member galaxies is completely dependent on the local gravitational potential \citep{Diaferio_1997} and shows a distinctive trumpet-like shape which is symmetric along the radial axis \citep{1987MNRAS.227....1K, 1993ApJ...418..544V, 1989AJ.....98..755R}, called the caustic amplitude, caustic surface or caustic profile $\mathcal{A}(r)$. The caustic amplitude then tracks the highest observable velocity ($v_{\rm{esc}}^{2}(r)$) at a given radius, which in turn is directly related to the gravitational potential, $\Phi(r)$,  of the cluster \citep{Diaferio_1997}. The gravitational potential can be written as
 
\begin{equation}
    \label{eqn:phi_caustic}
        -2\Phi(r)=\mathcal{A}^{2}(r)g(\beta),
    \end{equation}
where $g(\beta)$ is a function of the velocity anisotropy profile $\beta(r)= 1- \langle v_{\theta}^2 +v_{\phi}^2\rangle/2\langle v_{r}^2\rangle$ as 
\begin{equation}
    g(\beta)=\frac{3-2\beta(r)}{1-\beta(r)} \; .
    \label{eqn:gbeta}
\end{equation}
{Once the caustic surface is determined, one can estimate the mass profile either  remaining completely model independent \citep{Diaferio09} or  assuming a certain functional form that describes the gravitational potential \citep{Gifford13}, with appropriate assumptions of the velocity anisotropy $\beta(r)$.} We once again stress that the caustic method in the former approach is independent of the dynamical equilibrium state of the cluster, the specific form of $g(\beta)$, or the profile of the gravitational potential $\Phi(r)$. Instead, this technique captures a combined observable for $g(\beta)$ and $\Phi(r)$ through the measurement of the caustic amplitude $\mathcal{A}(r)$. The equation for estimating the  caustic mass profile of a spherical system can then be written as
\begin{equation}
        \label{eqn:Mass_Profile}
        \GN M(<r)=\int_{0}^{r}\mathcal{A}^{2}(r)\fbr dr
\end{equation}
where $\fbr =\fr g(\beta)$ and 
\begin{equation}
    \label{eqn:fofr}
    \mathcal{F}(r)=-2\pi \GN \frac{\rho(r)r^{2}}{\Phi(r)}\; ;
\end{equation}
$\rho(r)$ is the mass density profile of the spherical system, which we assume to be the NFW profile throughout the current analysis. \Cref{eqn:Mass_Profile} connects the mass profile to the density profile of a spherical system. In hierarchical clustering scenarios, the function $\mathcal{F}(r)$ shows minimal variations at different radial distances $r$ \citep{Diaferio_1997, Diaferio99}. Likewise, $\fbr$ changes slowly with $r$ because $\gbr\equiv g[\beta(r)]$ %the velocity anisotropy parameter $\beta$ i
is also a slowly varying function of $r$,
%\footnote{Needless to say, in the outermost regions of the cluster the anisotropy profile can show a larger deviation, while it can very well approximated as a constant within $\rf$.} 
as it has been validated in several works in varying ranges of redshifts, both from simulations and data \citep{Gifford_2013Ani, Svensmark:2019owr, Stark:2017vfa, Biviano:2021mtw}. Based on this property,  $\fbr$ can be replaced by a constant filling factor $\fb$ \citep{Diaferio99}.

A constant filling factor proves to be advantageous when it is applied in a practical way to individual clusters. It generally yields accurate escape velocity and mass profiles, although there can be occasional discrepancies from the actual profile, depending on the deviation from spherical symmetry and on the completeness of the member galaxies within a given survey {\citep[see for instance][]{Serra_2013}}. For example, if the observed galaxy sample has low completeness\footnote{Completeness refers to the number of galaxies in the sample with measured spectroscopic redshifts \citep{Sohn_2017}.} in the outer regions of the cluster, the caustic amplitude may be underestimated, leading to an underestimation of the mass. This event is particularly noticeable as the flattening of the mass profile when no galaxies are included in the sample (see also Sec. \ref{app:systematics}) and it should not be clearly inferred as an actual flattening of the physical mass profile of the cluster. 

The caustic method is notably important as an alternative to gravitational lensing  to measure the mass in the outer regions of a cluster \citep{Diaferio_2005}. Unlike lensing, it can be utilized for clusters at any redshift, as long as there are enough galaxies to sample the redshift diagram. 

%%%%%%%%%%%%%%%%%%%%%%%%%%%%%%%%%%%%%%%%%%
\section{Data} \label{sec:data}
%%%%%%%%%%%%%%%%%%%%%%%%%%%%%%%%%%%%%%%%%%

We compile our data from three different publicly available data sets, namely HeCS-SZ \citep{Rines_2016}, HeCS-redmapper \citep{Rines_2018} and HeCS \citep{Rines_2013}. HeCS-SZ is a MMT/Hectospec spectroscopic survey of 58 galaxy clusters with a total of 11,585 galaxies in the redshift range $z\in \{0.1, 0.3\}$. The Hectospec Cluster Survey of red-sequence-selected clusters (HeCS-red) includes 10,589 new or remeasured redshifts from MMT/Hectospec observations for 27 redMaPPer clusters with large estimated richness $\lambda > 64$\footnote{{The richness parameter $\lambda$ serves as an analog of the mass estimated through different techniques \citep{Rozo:2013vja, Simet:2016mzg, Sadibekova:2014goa}. In \cite{Rykoff:2007ub, Simet:2016mzg}, an analysis of stacked clusters with similar richness has been performed, which can serve as an alternate criteria for stacking, alongside velocity dispersion or mass (as implemented here).}}
in the redshift range $z \in \{0.08,0.25\}$. In \cite{Rines_2018}, a comparison of the optical richness to the dynamical masses has been performed for the HeCS-redMapper sample. The Hectospec Cluster Survey (HeCS) is a MMT/Hectospec spectroscopic survey of a flux-limited sample of X-ray-selected galaxy clusters in the redshift range $z \in \{0.1,0.3 \}$. \cite{Rines_2013} obtained 19,609 new redshifts for 58 clusters and measured velocity dispersions and mass profiles extending well into their infall regions. Furthermore, we take into account the HeCS-omnibus sample \citep{Sohn_2020}, which is a collection of all three datasets, to perform the comparisons with our sample and the necessary validations. 

We proceed with the estimation of the caustic masses for each cluster in these data sets. Firstly, we only take into account the galaxies for which the membership selection has already been performed and presented in their respective compilations. We then assign each galaxy to its corresponding cluster by finding $\theta$, its angular separation from the cluster center, and its redshift separation from the cluster center;  the minimum value of $\theta$ identifies the cluster to which the galaxy belongs. We limit the sample of galaxy members by only taking galaxies that are within $8\, \mathrm{h}^{-1} \Mpc$ and with $\Delta cz<5\times 10^3 \, \ks$ from the cluster center.  

{To obtain statistically relevant caustic profiles we only consider clusters with more than 40 galaxy members.\footnote{We could taking into account only clusters with more than 60 members. However, there are only four clusters with more than 40 and less than 60 galaxy members. Therefore, we proceed with including these four clusters in the analysis.} After these  selections, the HeCS-SZ sample is left with 48  clusters with a total of 10509 galaxies, the HeCS-red sample is left with 22 clusters with a total of 3463 galaxies, and the HeCS sample is reduced to 58 clusters with a total of 9721 galaxies.} The sample now contains  a total of 128 clusters; we further remove  6 clusters whose %due to the systematic\footnote{These clusters have 
relatively sparse redshift diagrams degrade our analysis. Two of these clusters are from the HeCS-red sample and the other four are from the HeCS-SZ sample.
%and introduce large spurious systematics to the stacked clusters, practically being outliers.} differences they contribute to the stacked analysis. 
To esimate the caustic masses, we construct the caustic profiles of the clusters \citep{Gifford13}\footnote{We use part of the code provided in \citep{Gifford13} to estimate the caustic surfaces. See also \cite{Kang:2024ets}, who have recently implemented a similar approach.} using the technique described in \citet{Diaferio99}. For consistency, in all the analyses and comparisons presented here %, of the caustic masses we compute, 
we adopt the filling factor $\mathcal{F}_{\beta}(r) = 0.5$. Our estimates of the caustic masses are in good agreement with those reported in the HeCS-SZ, HeCS-redMapper, and HeCS-omnibus samples.

%%%%%%%%%%%%%%%%%%%%%%%%%%%%%%%%%%%%%%%%%%
\section{Method} \label{sec:method}
%%%%%%%%%%%%%%%%%%%%%%%%%%%%%%%%%%%%%%%%%%

%%%%%%%%%%%%%%%%%%%%%%%%%%%%%%%%%%%%%%%%%%
\subsection{Stacked Clusters} \label{sec:Stacked_clusters}
%%%%%%%%%%%%%%%%%%%%%%%%%%%%%%%%%%%%%%%%%%
We employ the mass stacking method detailed in \citet{Gifford:2016plw} to obtain our stacked clusters. We bin our galaxy clusters in caustic mass so that each bin contains an approximately equal number of galaxies. By keeping the number of galaxies per bin similar, the width of the bins varies, reflecting the distribution of the galaxy cluster masses of our datasets. Given that low-mass clusters are more abundant in our dataset, the bins at the lower end of the mass spectrum are considerably narrower compared to those at the higher-mass end. We used our combined {dataset of 122} clusters with 23,639 galaxies to create 4 stacked clusters with $\sim 6000$ galaxies each. We anticipate that the large number of galaxies per stacked cluster will allow us to estimate robust caustic profiles, enhancing our analysis of these clusters.
Similarly to our approach in \cite{Butt:2024jes}, we utilize the cluster caustic profiles obtained from the velocity distributions of galaxies in redshift space and its relation with the gravitational potential in %the theoretical description of the caustic amplitude given by 
\cref{eqn:phi_caustic}. However, here we implement a procedure that is independent of the hydrostatic assumption to compute cluster masses; therefore, we constrain the modified gravity parameters using the caustic analysis alone. Specifically, we fit the potential $\Phi(r)$ of a NFW density profile (\cref{eqn:NFW_potential} below) to the profile derived from the caustic amplitude (\cref{eqn:phi_caustic}) adopting the velocity anisotropy parameter detailed in \cref{eq:betaT} below. %and \cref{eqn:NFW_potential} to the caustic amplitude obtained directly from the data.
For validation purposes, we also test our formalism on the cluster A2029  (App. \ref{sec:A2029}) and compare the constraints to those obtained when hydrostatic data are included \citep{Butt:2024jes}. 

Before proceeding to our analysis of  the caustic amplitudes of the stacked clusters, we compute the NFW masses of the individual clusters using the method described in \ref{sec:NFW Fitting} below, to perform various validations.  
We then proceed to estimate the caustic profiles of the stacked galaxy clusters and fit them to the NFW potential as done for the individual clusters. We also compare the masses of the stacked clusters with the median caustic masses and the {NFW-fitted caustic} masses of each mass bin. % and NFW masses of each mass range of stacked clusters. 
The caustic masses of the individual clusters forming the stacked clusters are estimated at the redshift of each cluster, with the caustic profiles computed using the cluster-centric projected radii and line-of-sight velocities. The NFW-fitted masses for the individual clusters are estimated independently of the concentration-mass (c-M) relation (see \cref{sec:cM}). For the stacked clusters, we measure the caustic masses by placing the cluster at $z=0.0$ and taking all of the composite clusters in their rest frames. When fitting the caustic mass profiles to the NFW profile we also utilize the prior of the c-M relation. The range over which we fit the caustic profile to the NFW one is $\rp < 4\, \Mpc$ for all the mass ranges. 

The advantage of computing the NFW mass using the caustic method is that the estimates are independent of the assumption on $\mathcal{F}_{\beta}(r)$. 
{However, assuming $\mathcal{F}_{\beta}=0.5$ for all the clusters, when computing the caustic mass, may yield possible discrepancies between the NFW-fitted masses and the caustic masses, because some clusters might not be properly described the NFW profile or might strongly deviate from spherical symmetry.}

\subsection{NFW Profile Fitting} \label{sec:NFW Fitting}

Under the assumption that the cluster is mostly dominated by dark matter and that it is spherically symmetric %, specifically in the range beyond 
at radii $\gtrsim 30-50 \, \mathrm{kpc}$,  we model the mass density using the NFW profile \citep{Schaller15, Hogan2017, Sartoris2020} %, which is given as ,
\begin{equation}
    \label{eqn:NFW_density_profile}
    \rho(r)=\frac{\rho_{s}}{(r/r_s)(1+r/r_s)^{2}}
\end{equation}
where $\rho_{s}$ is the characteristic density, and $r_{s}$ is the characteristic radius %, and the logarithmic slope $s= \diff \mathrm{ln}\rho/ \diff \mathrm{ln} r$ takes the isothermal value $s=-2$ 
\citep{Navarro:1995iw}. The corresponding cumulative mass profile is \citep{Cardone2020Aug,Haridasu:2021hzq} %in terms of $\Md$ and the concentration $\cd$ is given as,
\begin{equation}
    \label{eqn:NFW_mass_profile}
    M(<r)=\Md \frac{\mathrm{ln}(1+\cd x)-\cd x/(1+\cd x)}{\mathrm{ln}(1+\cd)-\cd/(1+\cd)}
\end{equation}
where $x=r/\Rd$, $\cd=\Rd/r_{s}$ is the concentration, and 
\begin{equation}
    \Md=\Delta \frac{4}{3}\pi\rho_{c}(z)\Rd^{3}.
\end{equation}
We  assume the value of $\Delta$ to be the usual value $\Delta = 200$ when making comparisons. The gravitational potential generated by the NFW density profile is 
\begin{equation}
    \label{eqn:NFW_potential}
    \Phi(r)= -\frac{4\pi G \rho_{s}r_{s}^{3}}{r}\mathrm{ln}(1+r/r_{s}).
\end{equation}

%We  assume the value of $\Delta$ to be the usual value $\Delta = 200$ when making comparisons. 
We utilize the same NFW mass profile to assess the masses in the GR scenario and chameleon screening\footnote{See \cite{Pizzuti:2024vjz, Pizzuti:2024stz} for a discussion on the effects of mass profile assumptions in constraining the chameleon screening model.} scenarios, as we have done previously in \citet{Boumechta:2023qhd} and \citet{Butt:2024jes}, respectively. 

According to \cref{eqn:phi_caustic}, we have $\mathcal{A}^{2}(r) = -2{\Phi(r)}/{\gbr}$, where $\Phi(r)$ is given by \cref{eqn:NFW_potential} and for $\gbr$ we have assumed %a fixed 
the velocity anisotropy profile %in the current analysis, which is 
parametrized with a generalized Tiret model (e.g. \citealt{Mamon19}). 
\begin{equation}
\label{eq:betaT}
\beta_\text{gT}(r)= \beta_{0} + (\beta_\infty -\beta_0)\frac{r}{r-r_\beta}\,
\end{equation} 
where $\beta_0$ and  $\beta_\infty $ represent the values of the anisotropy at the center and at large distances from the center, respectively; $r_\beta$ is the characteristic radius which we assume to be equivalent to the scale radius $r_\text{s}$ of the total mass profile,\footnote{Simulations show %It has been shown through simulations in \cite{Mamon:2004xk}, 
that the assumption of $r_\beta \sim r_{-2}$ can adequately model the velocity anisotropy profiles in galaxy clusters \citep{Mamon:2004xk}. {Here, $r_{-2}$ is defined as the radius at which the logarithmic slope of the density profile equals $-2$ (i.e., where ${\diff \ln\rho}/{\diff\ln r} = -2$).}} following %as was have earlier utilized in 
\citet{Pizzuti:2024stz} and \citet{ Mamon19}. \Cref{eq:betaT} provides a general description of possible anisotropy profiles in galaxy clusters, which is almost %also behaves as a 
constant in the outer regions of the galaxy cluster  \citep[see also \citealt{Gifford13}]{Diaferio99}. In general, the anisotropy of the stacked cluster is negligible when averaged over several clusters; we elaborate on this assumption in App.~\ref{app:systematics}. %However, we leave the full analysis for later work. 
In our analysis we set $\beta_0 = 0$, which reduces the generalized Tiret model to the Tiret model, and further fix $\beta_{\infty} = 0.5$, which remains fairly unconstrained as presented in App.~\ref{sec:A2029}. 

We fit the caustic profile obtained from the method described in \Cref{sec:Caustic} to the theoretical description of the caustic surface given in \cref{eqn:phi_caustic} using \cref{eqn:NFW_potential} and \cref{eq:betaT} to obtain the NFW-fitted caustic masses. The NFW-fitted caustic masses for the individual clusters are estimated while keeping $c_{200}$ fixed using the $c-M$ relation. 
For the caustic masses of the stacked clusters, in the GR case, the potential is
\begin{equation}
    \label{eqn:phi_GR}
    \Phi(r)=-\frac{GM(<r)}{r}-4\pi G \int_{r}^{\infty}\rho(x)xdx,
\end{equation}
while for the chameleon screening case, we use the potential given in \cref{eqn:grav_cham} below. 
In these potentials, $M_{200}$ and $c_{200}$ are fitting parameters. We choose to keep $c_{200}$ fixed in certain cases. As mentioned earlier, we fit over the range $\rp < 4\, \Mpc$ for the four stacked clusters. Furthermore, we bin the data of our caustic profile before fitting, thus reducing the weights of the data within $\rp = 0.5\, \Mpc$, where the dynamics are dominated by the Brightest Central Galaxy (BCG).

%%%%%%%%%%%%%%%%%%%%%%%%%%%%%%%%%%%%%%%%%%
\subsection{c-M relation} \label{sec:cM}
%%%%%%%%%%%%%%%%%%%%%%%%%%%%%%%%%%%%%%%%%%
We impose the concentration-mass ($c-M$) relation from \cite{Merten_2015} (hereafter we refer to this relation as M16) as a prior in our MCMC analysis. This relation is given by
\begin{equation}
\label{eqn:cM_martens}
    c_{200}(M_{200},z) = A \times \left(\frac{1.37}{1+z}\right)^{B} \times \left(\frac{M_{200}}{8\times10^{14} h^{-1} \mathrm{M}_{\odot} }\right)^{C}
\end{equation} 
where $A=3.66 \pm 0.16,$ $B=-0.14\pm 0.52$ and $C=-0.32 \pm 0.18$. 
We also consider other two concentration-mass relations from \cite{Dutton_2014}  \begin{equation}
    \label{eqn:dutton}
    \mathrm{log}_{10}c_{200} = 0.905-0.101\times\mathrm{log}_{10}\left(\frac{M_{200}}{10^{12}h^{-1}\mathrm{M}_{\odot}} \right)  
\end{equation}
and from \cite{Child_2018}
\begin{equation}
\label{eqn:cM_childs}
    c_{200}=A(1+z)^{d}M_{200}^{m}, 
\end{equation}
where $A=57.6$, $d= -0.37$ and $m=-0.078$; the latter relation has  a dispersion $\sigma_{c}=c_{200}/3$. We  validate that the $c-M$ relations estimated in \cite{Dutton_2014} and \cite{Child_2018} agree  %in very good agreement 
with each other, within the uncertainties, especially in the mass ranges of our analysis.\footnote{There exists a mild difference between the c-M relations at low masses $\Mf < 0.5 \, \times \Munit $, where \cref{eqn:dutton} \citep{Dutton_2014} predicts lower concentrations compared to \cref{eqn:cM_martens} \citep[M16]{Merten_2015}.}  We also find that the relation M16 is convenient to impose as a prior, dependent on mass and redshift, on the concentration values. Several other similar relations exist in the literature \citep{Meneghetti:2014xna, Ludlow:2016ifl, Bhattacharya2013}; they are substantially equivalent to each other for our purposes of setting a  prior on the $c-M$ relation.

%%%%%%%%%%%%%%%%%%%%%%%%%%%%%%%%%%%%%%%%%%
\subsection{Modified Gravity} \label{sec:Chameleon}
%%%%%%%%%%%%%%%%%%%%%%%%%%%%%%%%%%%%%%%%%%
Because we apply the caustic method to stacked clusters with data probing the outer regions of galaxy clusters, our analysis can  % Given our construction of the stacked caustic surfaces, the applications for such a dataset 
extend beyond the estimation of the mass of galaxy clusters in the standard scenarios. In this context, here we study a class of modified gravity models based on the chameleon field. The chameleon model \citep{Khoury:2003aq, Zaregonbadi:2022lpw, Tsujikawa:2009yf, Kraiselburd:2015vyf} modifies gravity by introducing a scalar field that is non-minimally coupled with the matter components, resulting in a fifth force that can be comparable in strength to the standard gravitational force. The chameleon mechanism is a two-parameter model: {i) $\beta$ determines the strength of the fifth force coupling to matter;  ii) $\phi_{\infty}$ gives the asymptotic value of the scalar field at large distances from the center of the cluster.} These parameters together provide the transition from the inner region, ruled by Newtonian dynamics, to the outer region where the fifth force becomes significant. In the absence of environmental effects, $\phi_{\infty}$ can be interpreted as the cosmological background value of the chameleon field. 

The chameleon mechanism operates whenever a scalar field is coupled to matter in such a way that its effective mass depends on the local matter density.
The force mediated by this scalar field between matter particles can be of gravitational strength; however, its range decreases as the ambient matter density increases, which allows it to remain undetected in high-density regions. In regions where the mass density is low,  the scalar field is light and can mediate a fifth force comparable to the gravitational force. Conversely, near Earth, where experimental measurements are conducted and local density is high, the scalar field acquires a large mass, resulting in short-range effects that make it unobservable. {This behavior is achieved with a canonical scalar field with suitable self-interaction potential $V(\phi)$, which is a decreasing function of $\phi$ \citep{Khoury:2003aq}. For instance, we assume a usual power-law potential of the form $V(\phi) = \Lambda^{n+4}\phi^{-n}$. However, the scalar field constrained within the cluster is not sensitive to the parameters $\{\Lambda,\, n\}$ \citep{Terukina:2013eqa}.} The part of the Lagrangian involving the chameleon scalar field, in the weak-field limit and for the non-relativistic matter, is \citep{Terukina:2013eqa, khoury2013les}
\begin{equation}
    L_{\mathrm{chameleon}}=-\frac{1}{2}(\partial\phi)^{2}-V(\phi)-\frac{g\phi}{M_{\mathrm{Pl}}}\rho_{\mathrm{m}}
\end{equation}
where $M_{\mathrm{Pl}}$ is the Planck mass.
The dimensionless coupling parameter $g$ is assumed to be of order $O(1)$, corresponding to the gravitational strength coupling. In regions of high density, the mass of the chameleon field increases, leading to a decrease in the range of interaction. Thus, the effects of the fifth force are screened, and GR is recovered. In low-density environments, the fifth force is not screened and the effects of the chameleon field can be observed. 
The modified gravitational potential $\Phi$ under this mechanism is \citep{Terukina:2013eqa, Wilcox:2015kna}:
\begin{equation}
    \label{eqn:grav_cham}
    \frac{ \diff { \Phi(r)}}{\diff {r}}=\frac{G_{\mathrm{N}}{M}(r)}{r^{2}} + \beta \frac{\diff{\phi}}{\diff {r}}\, .
\end{equation}

The chameleon field $\phi$ mediates a long-range fifth force when the matter density remains significantly larger than the background density, and the scalar field has not yet settled into the minimum of its effective potential. The chameleon field becomes effective beyond a critical radius, $r_c$, below which the field is completely screened. This radius is determined by the relation \citep{Terukina:2013eqa}:
\begin{equation}
    \label{eqn:critical_radius}
    1+\frac{r_c}{r_s}=\frac{\beta\rho_s r_s^2}{M_{\mathrm{Pl}}\phi_{\infty}},
\end{equation}
where $\rho_{s}$ and $r_{s}$ are the characteristic density and characteristic radius, respectively. 

%%%%%%%%%%%%%%%%%%%%%%%%%%%%%%%%%%%%%%%%%%
\subsection{Analysis} \label{sec:analysis}
%%%%%%%%%%%%%%%%%%%%%%%%%%%%%%%%%%%%%%%%%%

To perform the Bayesian analysis, we implement a simple Gaussian likelihood for $N$ number of data points of the caustic surface:

\begin{equation}
\label{eqn:fitlikelihood}
\begin{split}
    -2 \ln\left[\mathcal{L}_{\rm cau}(\Theta)\right] &= \sum_{i}^{N}\frac{\left(A_i - \sqrt\frac{-\Phi_{\rm NFW}(r_i, \Theta)}{g_{\beta}(r_i, \Theta)}\right)^{2}}{\sigma_{A_i}^2 + \sigma_{\rm int}^2} \\ &+ N \ln[2 \pi (\sigma_{A_i}^2 + \sigma_{\rm int}^2)] 
\end{split}
\end{equation}

where $\Theta$ is the set of parameters $\{\Mf, \cf, \beta, \phi_{\infty}\}$, $A_i$ is the caustic amplitude at radius $r_i$, and $\sigma_{A_i}$ is the error on the caustic amplitude. { Here, $\Phi_{\rm NFW}$ is the NFW potential and $g_{\beta}$ is a function of the velocity anisotropy profile in \cref{eq:betaT}. Note that here we have rewritten the potential $\Phi_{\rm NFW}$ in \cref{eqn:NFW_potential} as a function of $\Mf,\, \cf$}. We also include $\sigma_{\rm int}$ as the intrinsic scatter parameter in the fit of the caustic amplitude. Here, we have constructed the data set so that the caustic surface is estimated with a resolution of $\Delta r_i = 0.01 \, \Mpc$ which is then reduced to a resolution of $\Delta r_i = 0.1 \, \Mpc$ implying 40 data points in the range $\rp \leq 4\, \Mpc$ for fitting purposes. The final error on the reduced data points is constructed as the weighted average of the errors on the original data points, also ensuring complete independence of the data in the different radial bins. This provides a balance between the computational time of the MCMC and the number of data points, providing the penalty term,\footnote{We have tested our likelihood with different numbers of reduced data points, finding good agreement for a number in the range 30-60. %However, we find that 
With the high resolution of the bins with an arbitrarily large number of points, one can provide extremely tight constraints because of the domination of the penalty term over the $\chi^2$ term.} which is the second term on the r.h.s of \cref{eqn:fitlikelihood}. Note that when performing the validations for the systematics as presented in App. \ref{app:beta}, the free parameters appropriately include %other parameters such as 
$\{\beta_0$ and $\beta_{\infty}\}$ of the velocity anisotropy profile. 

We also include the c-M relation, \cref{eqn:cM_martens} as a prior in our analysis; for this relation we use the uncertainties %reported on the fitting formula 
reported in \cite{Merten_2015}. {We then add to the above likelihood 
$\mathcal{L}_{\rm cau}$, which fits the caustic surface, the Gaussian prior 
\begin{align*}
    \label{eqn:prior}
    -2 \ln\left[\mathcal{\mathcal{L}_{\rm cM}}(\Mf)\right] &= \left(\frac{\cf - \cf^{\rm M16}(\Mf)}{\sigma_{\cf}(\Mf)}\right)^{2} \\ &+ \ln(2\pi \sigma_{\cf}^2)
\end{align*}
where $\cf^{\rm M16}(\Mf)$ is the concentration-mass fitting formula, i.e, the M16 relation, and $\sigma_{\cf}(\Mf)$ is the uncertainty in the relation, which is also a function of the mass, as depicted by the gray region in \Cref{fig:GR_contours}. Finally, the joint likelihood when including the $c-M$ prior is simply given as $\mathcal{L}_{\rm joint} = \mathcal{L}_{\rm cau} +\mathcal{L}_{\rm cM} $.} 

We perform a simple Bayesian analysis using the publicly available \texttt{EMCEE}\footnote{Available at: \href{http://dfm.io/emcee/current/}{http://dfm.io/emcee/current/}} package \citep{Foreman-Mackey13} to perform the sampling and \texttt{GetDist}\footnote{Available at: \href{https://getdist.readthedocs.io/}{https://getdist.readthedocs.io/}} \citep{Lewis:2019xzd} and \texttt{corner}\footnote{Available at: \href{https://corner.readthedocs.io/en/latest/}{https://corner.readthedocs.io/en/latest/}} \citep{corner} to analyze the chains. We impose sufficiently large uniform priors on all the free parameters and running longer chains to ensure convergence, in the analysis. We then perform an importance sampling analysis of the posteriors obtained for the 4 mass ranges on the modified gravity parameters, to present a final constraint.

\begin{figure*}
  \centering
  \includegraphics[width=0.50\linewidth]{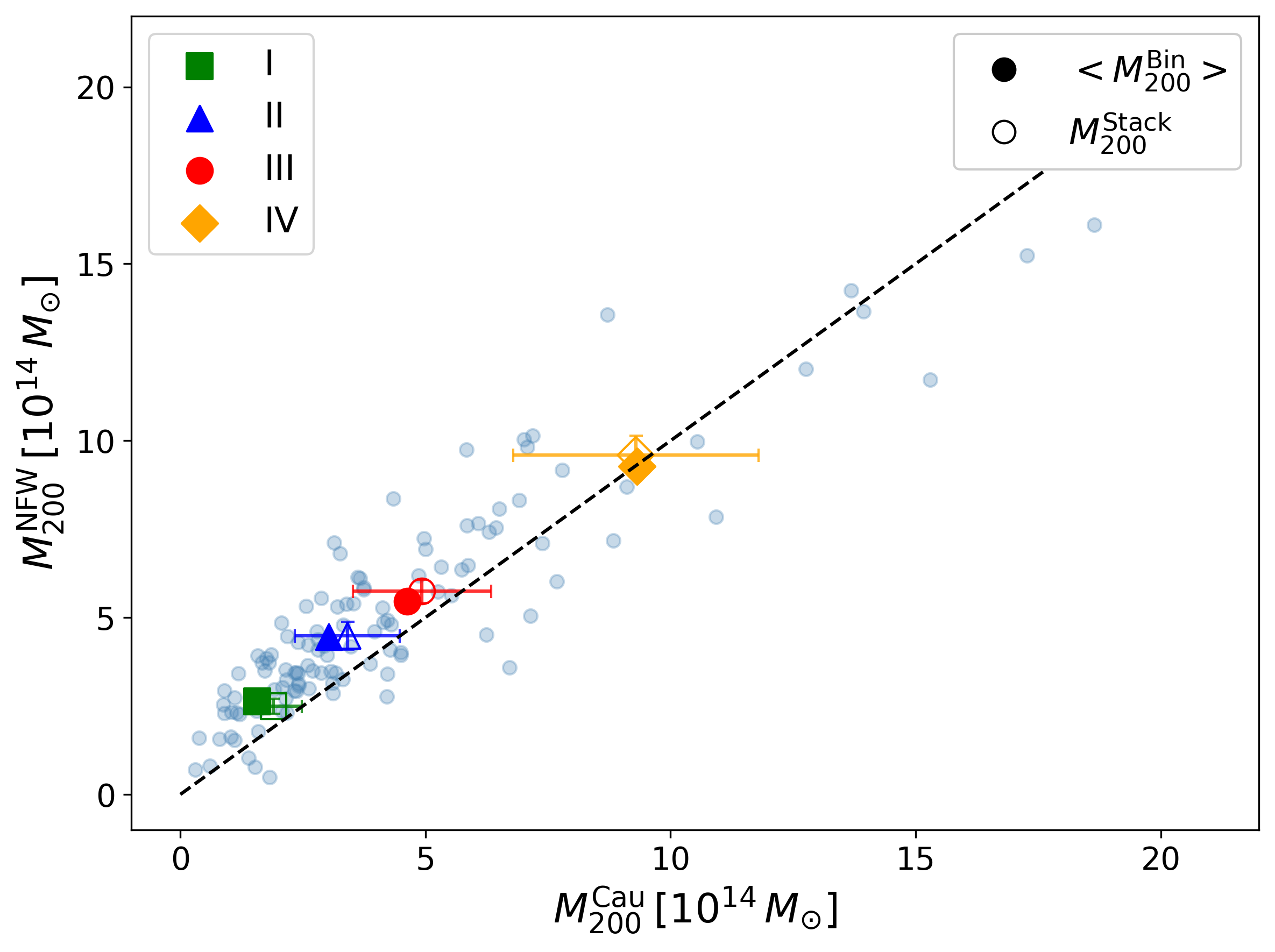}
  \includegraphics[width=0.49\linewidth]{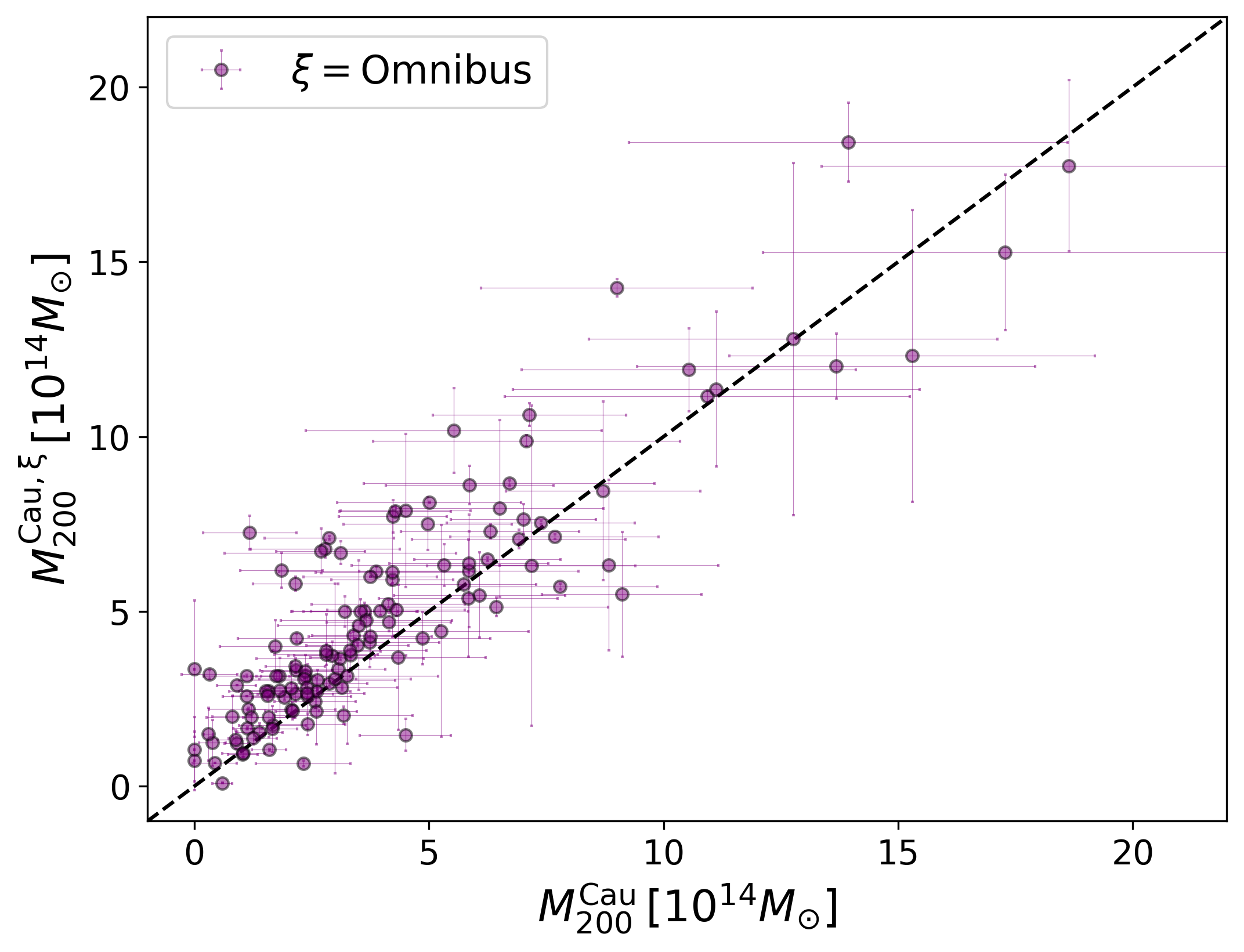}
  \caption{\textit{Left}: Comparison of the caustic ($\Mcau$) with the NFW ($\Mnfw$) masses (see \cref{sec:Stacked_clusters} for details). The fitting range for the NFW masses is $0<(\rp /\mathrm {Mpc})<4$. The open symbols represent the masses of the stacked clusters and the solid symbols represent the mean masses of the stacked clusters. The small filled circles show the individual clusters. \textit{Right}: We compare the caustic masses estimated in our analysis with those presented in the earlier analysis of the HeCS-omnibus catalog \citep{Sohn_2020}. }
  \label{fig:Mnfw_Mcau} 
\end{figure*}

\begin{figure*}[!ht]
\centering
\includegraphics[width=0.495\textwidth]{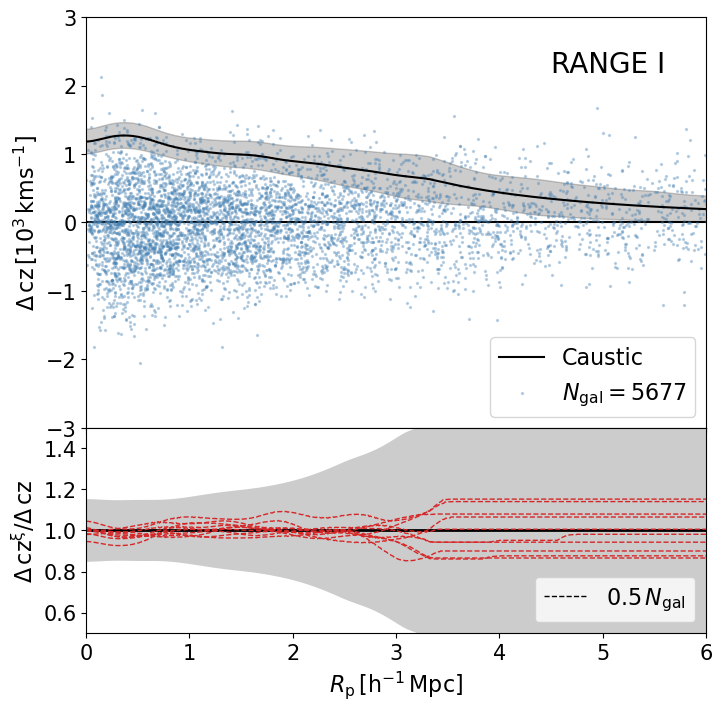} 
\includegraphics[width=0.495\textwidth]{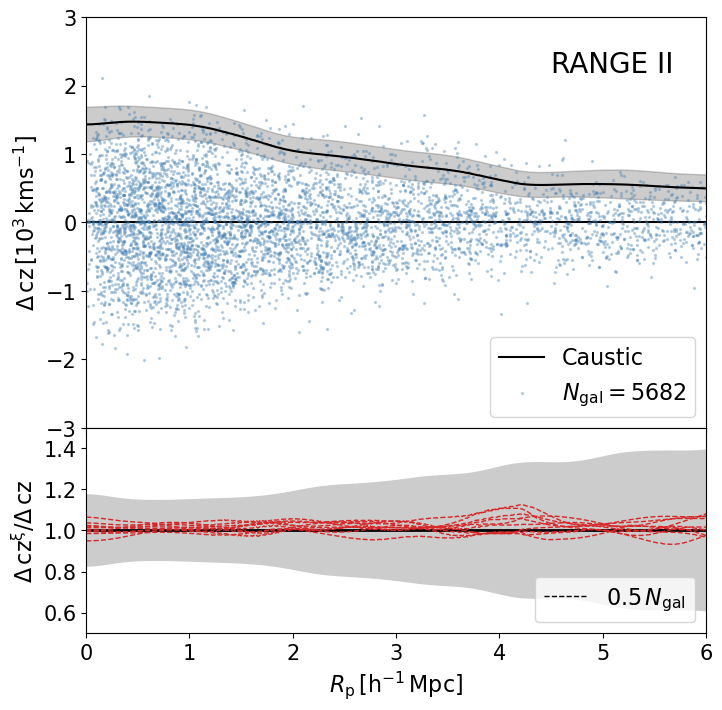} 
\includegraphics[width=0.495\textwidth]{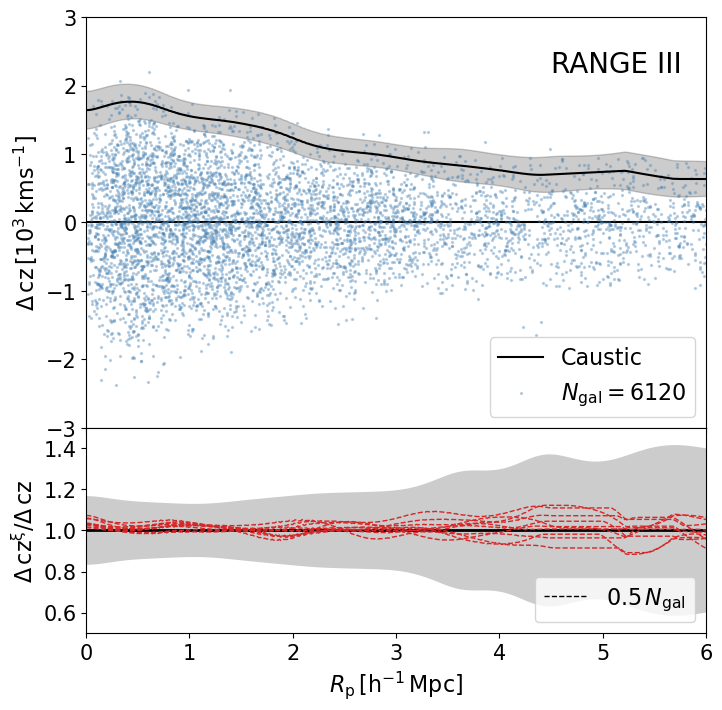} 
\includegraphics[width=0.495\textwidth]{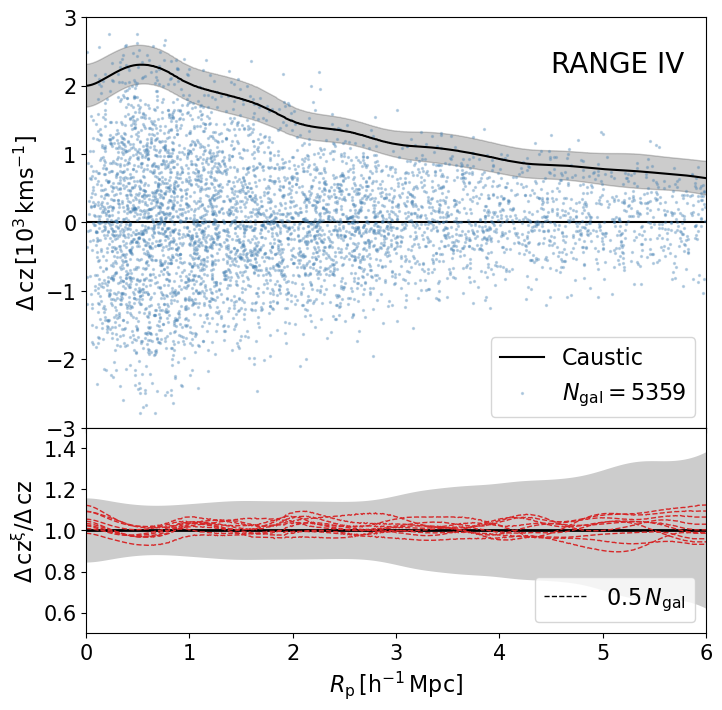} 
\caption{ \textit{Top Panels}: Distribution of the galaxies (blue points) in the redshift diagram, namely the plane of the redshift vs. the projected distance from the cluster center, of the four stacked clusters. The solid line and the gray area show the caustic location along with its uncertainty. The lower caustic is omitted.
\textit{Bottom Panels}: Ratio of the mean caustic amplitude obtained from the bootstrap analysis containing only $50\%$ of the total galaxies and the caustic profile obtained with the full dataset. The shaded region represents the error bar of the caustic amplitude for the full dataset. }
\label{fig:caustic_surface}
\end{figure*}

%%%%%%%%%%%%%%%%%%%%%%%%%%%%%%%%%%%%%%%%%%
\section{Results and Discussion} \label{sec:results}
%%%%%%%%%%%%%%%%%%%%%%%%%%%%%%%%%%%%%%%%%% 
In this section, we present the results obtained from our analysis starting with the comparison of the caustic masses of the individual clusters to their NFW masses. Following this, we discuss the caustic profiles of the stacked clusters and compare the caustic masses with the NFW masses of the stacked clusters.   

\subsection{Mass comparison of individual clusters} 
\label{sec:mass_comp} 
To remain consistent within our analysis of the estimate of the caustic masses of the individual clusters and, later, of the stacked clusters, we first begin by estimating the caustic masses of all individual clusters assuming $\fbr = 0.5$. As shown in the right panel of \cref{fig:Mnfw_Mcau}, we find results consistent with the caustic masses compiled in the original HeCS-omnibus sample \citep{Sohn_2020}. The dispersion in the difference between their estimates and ours is of the order of $\sim 10\%$. The discrepant masses are due to the differences in the identification of the caustic surfaces in the analyses and are within the uncertainties. %However, hereafter we are interested in the relative comparison of the caustic masses to the NFW masses. Given the agreement between the masses estimated in our formalism and in the literature, we 

We now proceed to make a comparison of the model-independent caustic masses with the NFW-based masses estimated through the reconstructed caustic surface, as elaborated in \cref{sec:method}. In the left panel of \cref{fig:Mnfw_Mcau}, we show the comparison between these different methods, which illustrates very good agreement, within a {%acceptable 
scatter $\sim 10\%$}. {Please note that here we have not yet utilized the c-M relation as a prior, and we find that the concentration parameter, which gives the slope of the density profile, is not tightly constrained. This loose constraint originates from the fact that the concentration parameter probes %is clearly because this parameter constrains 
the transition of the slope of the density profile at the scale radius $r_{\rm s}$, % of the clusters, 
which is well within the virial radius, $R_{200}$; estimating $r_{\rm s}$ requires a more accurate estimation of the caustic surface in this range, where the caustic method is not expected to be accurate by construction \citep{Diaferio99}.} % for each individual cluster.} 
We then utilize the c-M relation as a prior, fixing the concentration of individual galaxy clusters following 
\Cref{eqn:cM_martens}. We find good agreement with the caustic masses, with a mildly reduced dispersion compared to the case without the c-M prior. %to that in the former case. 
Indeed, as we present in  \Cref{sec:BA} below, the inclusion of the c-M prior gives only a mild difference in the analysis of the stacked clusters.
 
{In essence, we show that either assuming the NFW density profile (both with and without the imposition of the c-M relation) or assuming $\fbr = 0.5$ returns consistent results. This finding validates the two techniques simultaneously, as it was originally suggested in \citet{Diaferio_1997}. We intend to elaborate on this interesting result in a dedicated {future} work.}

{\renewcommand{\arraystretch}{1.7}
\setlength{\tabcolsep}{4.5pt}

\begin{table*}[!ht] 
    \caption{Constraints on the mass in the case of GR, shown as $68\%$ C.L. limits for each of the four stacked mass ranges as described by the first 5 columns.  }
    \label{tab:GR_Masses}
    \centering
    \begin{tabular}{ccccccccccccc}
        \hline
        Range & \(M^{\rm Bin}_{200}\) & \(\Ngal\) & \(\Ngal ^{\rm{r}_{p}<4.0}\) & \(\Nclus\)   & \(\left<M^{\rm Bin}_{200} \right>\)  &   \(\Mcau (\fb = 0.5)\)& c-M prior & \(\cnfw\) & \(\Mnfw\) & \(\log(\sigma_{\rm int})\)\\
        & \([\Munit]\)& &  & &\([\Munit]\) & \( [\Munit]\) & &   &\([\Munit]\) &   &  \\

        \hline
        \hline
        \multirow{2}{*}{I} & \multirow{2}{*}{$0.30-2.40$} & \multirow{2}{*}{$5677$}& \multirow{2}{*}{$4839$} & \multirow{2}{*}{$42$}& \multirow{2}{*}{$1.57_{-0.11}^{+0.11}$} &  \multirow{2}{*}{$1.90_{-0.58}^{+0.58}$} & with & $3.75_{-0.71}^{+0.80}$ & $2.49_{-0.23}^{+0.20}$ & $-3.36^{+0.59}_{-1.50}$ &  \\
        & & & & & & & w/o& $2.79^{+0.67}_{-0.55}$& $2.19^{+0.24}_{-0.25}$ & $-3.41^{+0.62}_{-1.52}$ & \\
        \multirow{2}{*}{II} & \multirow{2}{*}{$2.40-3.74$}& \multirow{2}{*}{$5682$} &\multirow{2}{*}{$4411$} & \multirow{2}{*}{$31$}& \multirow{2}{*}{$3.03_{-0.24}^{+0.24}$} & \multirow{2}{*}{ $3.41_{-1.07}^{+1.07}$} & with &$4.24_{-0.57}^{+0.57}$  & $4.48_{-0.29}^{+0.25}$ &$-2.68^{+0.24}_{-0.30}$ & \\
        & & & & & & & w/o & $3.48^{+0.89}_{-0.75}$ &$4.14^{+0.42}_{-0.47}$ & $-2.60^{+0.21}_{-0.26}$\\
        \multirow{2}{*}{III} & \multirow{2}{*}{$3.74-5.85$} & \multirow{2}{*}{$6120$}& \multirow{2}{*}{$4795$} & \multirow{2}{*}{$23$} & \multirow{2}{*}{$4.63_{-0.36}^{+0.36}$} & \multirow{2}{*}{$4.93_{-1.41}^{+1.41}$} & with & $4.35_{-0.49}^{+0.48}$  & $5.74_{-0.33}^{+0.32}$ &$-2.15^{+0.17}_{-0.18}$ & \\
        & & & & & & & w/o & $4.48^{+1.24}_{-1.02}$ & $5.78^{+0.53}_{-0.56}$& $-2.14^{+0.17}_{-0.18}$ &\\
        \multirow{2}{*}{IV} & \multirow{2}{*}{$5.85-18.64$} & \multirow{2}{*}{$5359$} & \multirow{2}{*}{$4466$} & \multirow{2}{*}{$26$} &\multirow{2}{*}{$9.31_{-0.60}^{+0.60}$} & \multirow{2}{*}{$9.29_{-2.50}^{+2.50}$} & with & $3.91_{-0.39}^{+0.37}$ & $9.59_{-0.53}^{+0.56}$ &$-1.70^{+0.14}_{-0.14}$ &
        \\
        & & & & & & & w/o & $5.47^{+1.47}_{-1.25}$ & $10.65^{+0.85}_{-0.92}$& $-1.75^{+0.15}_{-0.15}$&\\

        \hline 
    \end{tabular}
\end{table*}
}

\begin{figure*}[!ht]
    \includegraphics[scale = 0.61]{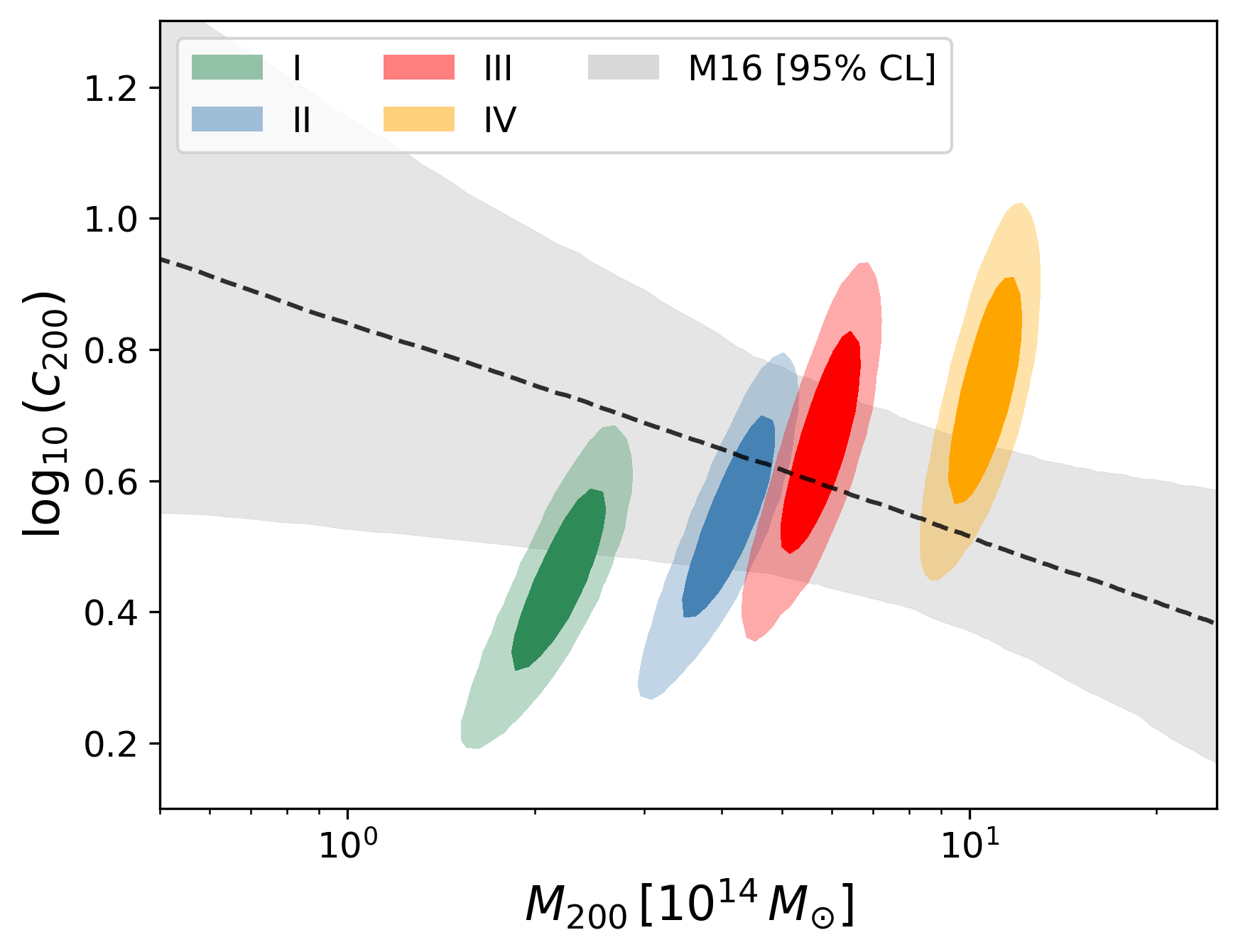}
    \includegraphics[scale = 0.61]{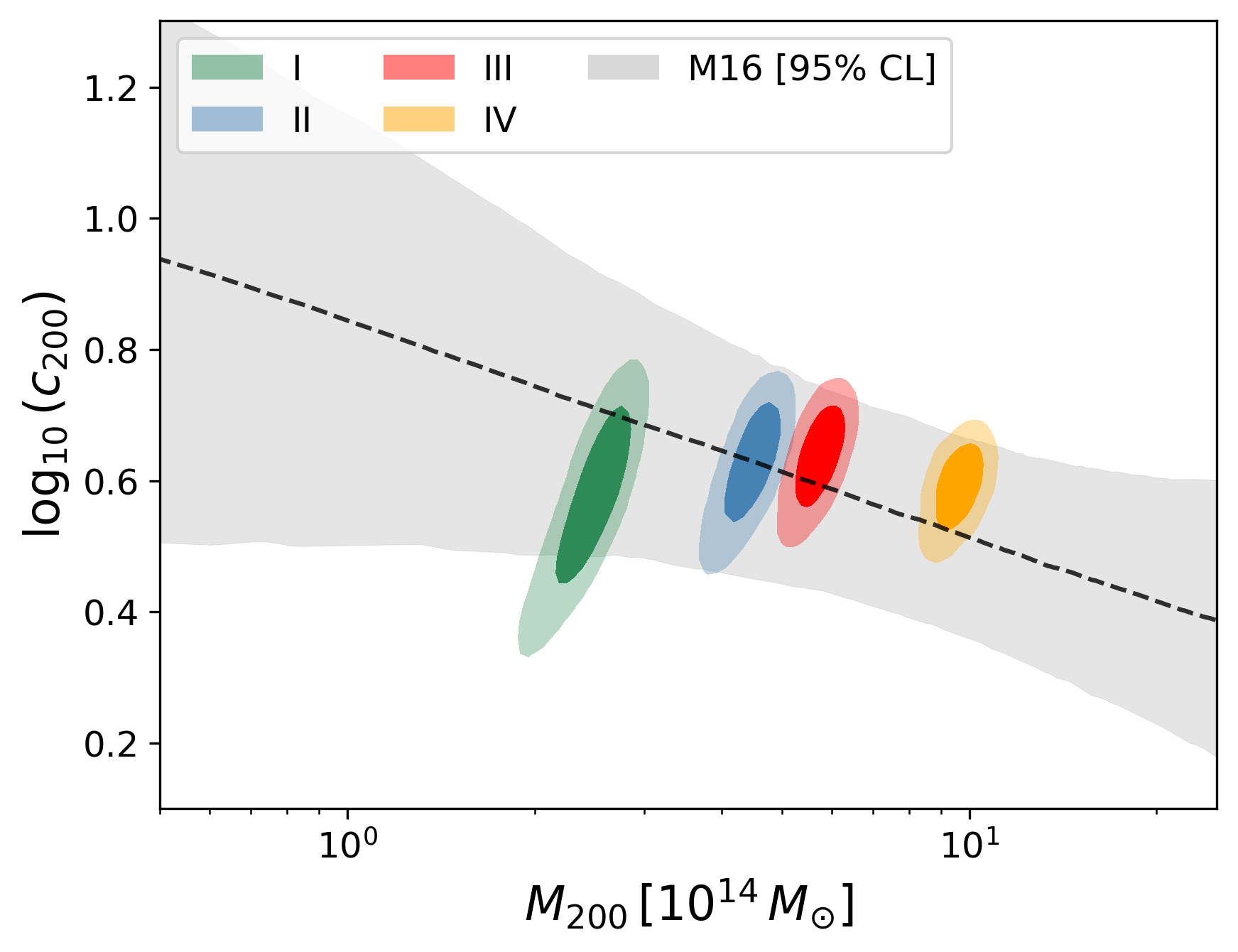}
  \caption{\textit{Left}: We show the posteriors for the $68\%$ and $95\%$ C.L. for the mass and concentration parameters in the standard GR case from the NFW mass profile for each of the mass ranges as described in \cref{tab:GR_Masses}. \textit{Right}: Same as the left panel but with the c-M relation M16 \citep{Merten_2015} imposed as a prior. We show the $95\%$ C.L. limits of the M16 in gray.}
  \label{fig:GR_contours} 
\end{figure*}

\subsection{Caustic location of the stacked clusters}
\label{sec:Caustic_profile}

Following the methodology described in \cref{sec:Stacked_clusters}, we now locate the caustics for the 4 stacked clusters. The second column of \cref{tab:GR_Masses} gives the upper and lower limits of the ranges of the mass bins. As shown in \cref{fig:caustic_surface}, we obtain a very well-estimated mean and uncertainty of the caustic location. %Although the radial range available to construct the caustic surface extends to the range of the caustic phase space itself (restricted to $8\, \hmpc$), t
The uncertainty of the caustic location is limited by the number of galaxies present within a given radial annulus, as seen in the lower panels of \cref{fig:caustic_surface}, and increases towards the outer regions. In the lower panels of \cref{fig:caustic_surface}, we show a few bootstrap cross-validations (red-dashed lines) of the caustic locations obtained by randomly selecting $50\%$ of the galaxies in the dataset. We find that the mean caustic location obtained with the full dataset is well within the dispersion of several bootstrapped caustic locations, which helps us validate our location of the caustics. % the completeness of the caustic surface we have constructed. 
Specifically, for stacked cluster I, the uncertainty on the caustic location increases %is not well estimated 
in the outer regions $\rp \gtrsim 3.5\, \Mpc$, where the redshift diagram is more sparsely sampled. %is due to the lack of sufficient galaxies in the redshift space. 
At any rate, we limit our analysis to the %However, we remain with 
the radial range of $0 \leq \rp \leq 4 \, [\Mpc]$\footnote{We utilize the same radial range also for the stacked cluster I, as the large uncertainty in the determination of the caustic location beyond $\rp \geq 3 \, [\Mpc]$ leaves %allows 
the overall fit unaffected. } 
%to be unbiased.} 
as our primary dataset for all the four stacked clusters, chosen to avoid any possible bias. Ideally, a completeness of $> 95\%$ of the galaxy sample would be necessary to accurately locate the caustics \citep{Serra_2013}. Different tests %We perform several tests estimating the masses, 
utilizing different radial ranges are %as 
presented %, and elaborate on the {same} 
in App.~\ref{app:systematics}. 

We now perform the estimation of the caustic and  NFW masses of the stacked clusters.
In the left panel of \cref{fig:Mnfw_Mcau}, we show the mean mass of each of the mass bins we construct for the stacked clusters. We find that the mean caustic masses of the stacked clusters are in good agreement with the mean NFW masses, as shown by the filled markers for each mass range. Similarly, we also find that the caustic mass of the stacked cluster is in excellent agreement with the best-fit NFW  estimated mass of the stacked cluster. This is an important validation of our method, as it shows that the stacked clusters are representative of individual clusters, which allows us to proceed with the subsequent analysis.

\begin{figure}
    % \centering
    \includegraphics[width=0.47\textwidth]{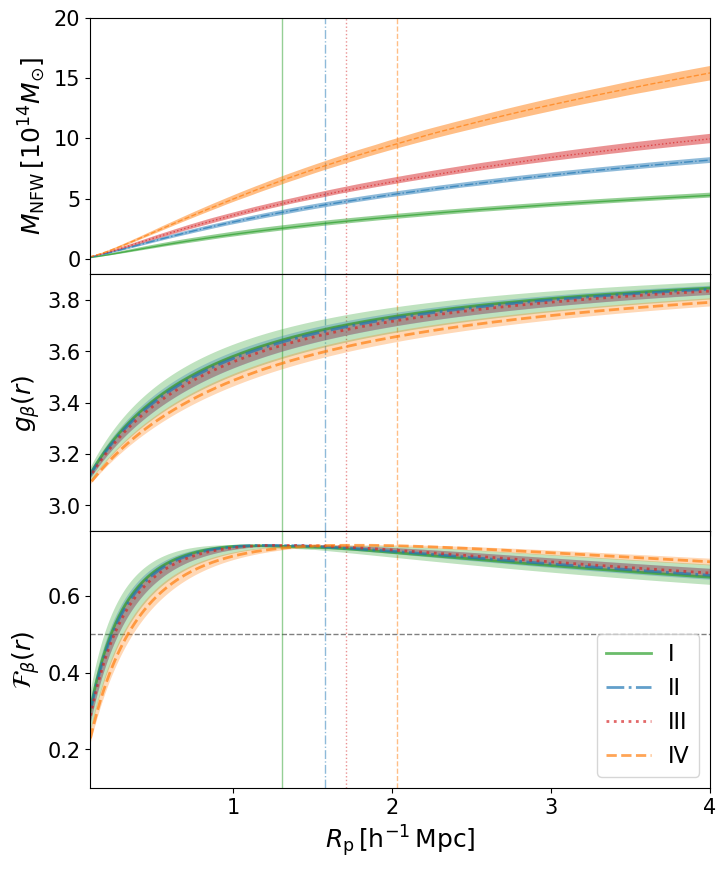}
    \caption{Mean and $1\sigma$ C.L. regions of $M_{\mathrm{NFW}}(r)$ (\textit{top} panel) profiles for the case of GR for each of the 4 ranges. The vertical lines %in the middle panel 
    represent $r_{200}$ for each of the 4 ranges. Similarly, of $g_{\beta}(r)$ (\textit{center} panel) and $\fbr$ (\textit{bottom} panel) for the case of GR for each of the 4 ranges. The horizontal dotted line in the bottom panel represents $\fbr = 0.5$. Here all the profiles are obtained from the analysis including the c-M prior.}
    \label{fig:fbr}
\end{figure}

\subsection{Constraints from the Bayesian Analysis}
\label{sec:BA}

In \cref{fig:GR_contours}, we show the posterior contours for the mass and concentration parameters in the standard GR case, fitted to the NFW mass profile, for each of the mass ranges presented in \cref{tab:GR_Masses}. The constraints for the mass and concentration parameters are presented in \cref{tab:GR_Masses}. As can be seen, we find good agreement for the mass estimates in each mass range compared to the model-independent caustic expectation. In the left panel of \cref{fig:GR_contours}, we show the posteriors without the inclusion of the c-M relation of \cref{eqn:cM_martens}, which are completely consistent within the uncertainty of the c-M relation. We find that the concentration parameter is not well constrained with larger uncertainties in this case, which is expected, as the slope of the density profile is better adapted to clusters within the virial range $R_{200}$, and requires better estimation of the caustic surface for each of the individual clusters. We then include the c-M relation (\Cref{eqn:cM_martens}) as a prior, utilizing also the uncertainty in the relation, where we find a very good agreement with the caustic masses, with a similar uncertainty in the mass estimates to the former case and significant reduction in the error on the concentration parameter.

\begin{figure*}[!ht]
    \includegraphics[scale = 0.48]{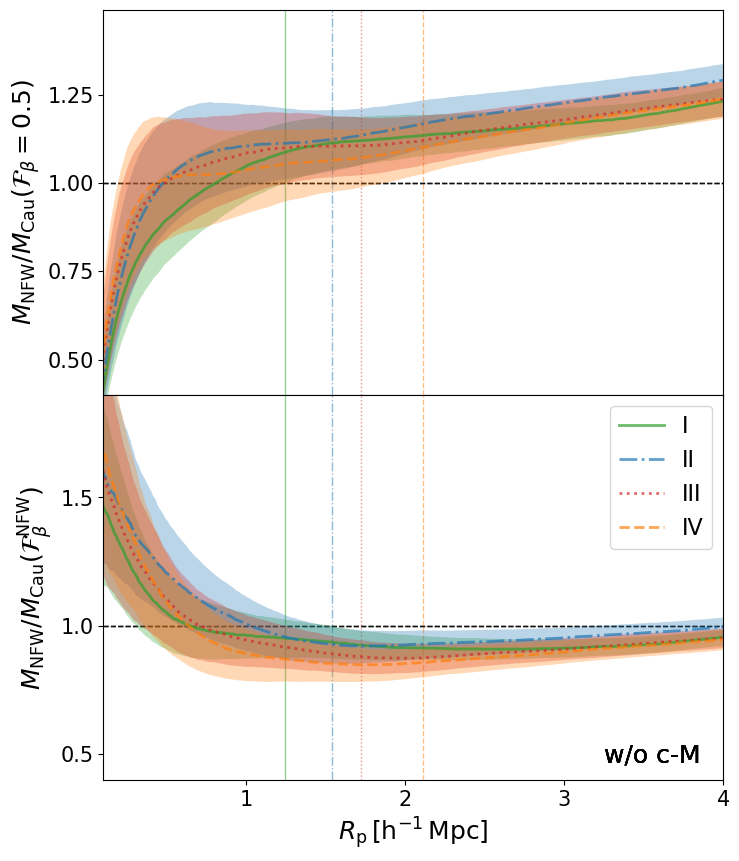}
    \includegraphics[scale = 0.48]{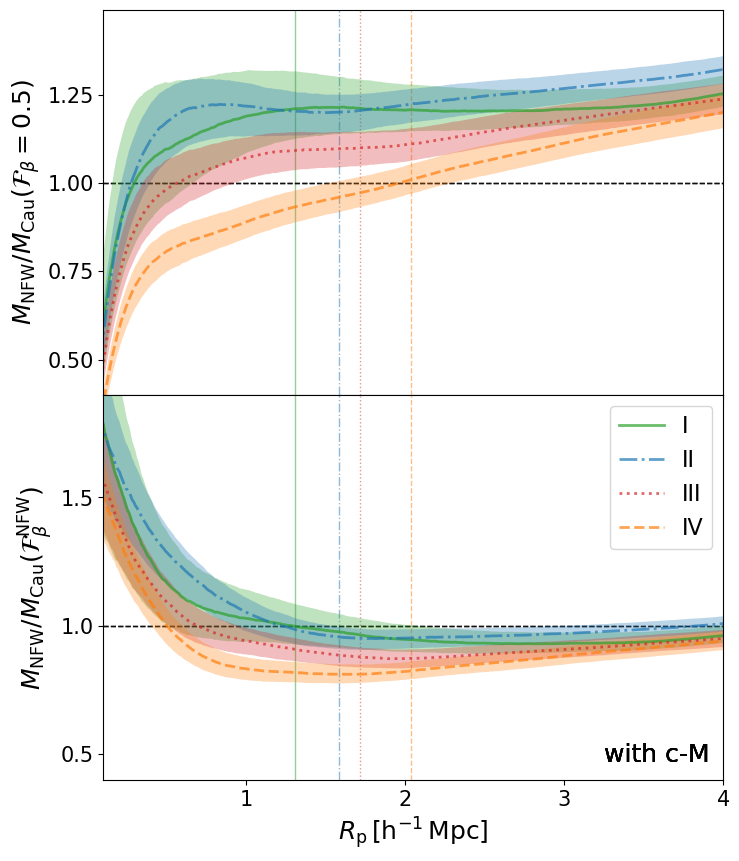}
  \caption{\textit{Left}: {Comparison of $M_{\mathrm{NFW}}(r)/M_{\mathrm{Cau}}(r)$ (bottom panel) for the case of General Relativity for each of the 4 ranges when the caustic mass is estimated assuming $\fbr$ computed from the NFW fit to the caustic surface. In contrast, the top panel shows the same comparison when the $\fbr = 0.5$ is amused to estimate the caustic mass. The vertical lines in the middle panel represent $r_{200}$ for each of the 4 ranges. \textit{Right}: Same as the \textit{Left} panel but with the inclusion of the c-M prior.} }
  \label{fig:fbr2}
\end{figure*}

In \cref{fig:fbr}, we show the mass profiles and the estimated variation in $\gbr$ and $\fbr$ for the 4 ranges. Firstly, we note an extremely good agreement for the estimation of $\fbr$ across all the 4 mass ranges showing an almost constant-like behavior with a slow decline as we go further from the cluster center, with the expected steep increase in the inner-most regions. This is an independent validation of the caustic technique to estimate the masses of galaxy clusters when the calibration factor $\fbr$ is well estimated. While we find this to be a very good assessment, we estimate { $\fb(\rp \sim \rf) \sim 0.65 $} around the virial radius of the cluster; in the outskirts of the cluster $\fb(\rp)$ mildly increases, in agreement with the usual range of $0.5\leq \fbr\leq 0.7$ presented in literature \citep{Diaferio_2005, Diaferio09, Gifford:2016plw}. 

However, note that this estimation of $\fbr$ is limited by the assumption of a fixed anisotropy profile. As shown in \cref{fig:fbr}, the variation in $\gbr$ is completely driven by the uncertainty of the scale radius, $r_{s}$, of the NFW profile, which is assumed to coincide with $r_{\beta}$ of the Tiret's profile in \cref{eq:betaT}. {While the values of $\beta(r)$ are expected to vary in the range  $-\infty \leq \beta \leq1$, $\gbr$ varies in a much more limited range than $\beta(r)$. Also, as we show in App \ref{sec:A2029}, validating our formalism using the cluster A2029, the imposition of the c-M relation as a prior allows us to constrain the mass and concentration parameters better even if the anisotropy parameters are loosely constrained. When we perform the same exercise with our stacked clusters, marginalizing over the anisotropy parameters, we find that the imposition of the c-M relation provides good agreement with the constraints with mildly enlarged uncertainties. Heuristically, this immense advantage of using the c-M relation can be understood by contrasting the orthogonal correlations of the $\{\cf, \Mf\}$ posteriors of our MCMC analysis and the c-M relation as shown in \cref{fig:GR_contours}. In any case, to completely take advantage of the previously mentioned opportunity to validate/falsify the $\LCDM$ model scenario, one would need to perform a detailed assessment, also estimating the anisotropy, by solving the Jeans equation \citep{Falco:2013zya, Svensmark:2019owr}, simultaneously. We leave this to a {future} dedicated analysis.}

{\renewcommand{\arraystretch}{1.5}
\setlength{\tabcolsep}{8pt}
\begin{table}[h]
  \centering  \begin{tabular}{ccc}
    \hline
    Range & \multicolumn{2}{c}{$\fbr$} \\
             & w/o c-M & with c-M \\ \hline 
             \hline
    I   & $0.58^{+0.11}_{-0.11}$ & $0.66^{+0.09}_{-0.09}$            \\ 
    II  & $0.61^{+0.11}_{-0.11}$  & $0.65^{+0.06}_{-0.06}$           \\ 
    III   & $0.59^{+0.09}_{-0.09}$  & $0.58^{+0.06}_{-0.06}$          \\ 
    IV    & $0.57^{+0.08}_{-0.08}$   & $0.52^{+0.06}_{-0.06}$        \\ \hline
    Joint & $0.59^{+0.05}_{-0.05}$ & -- \\ \hline
  \end{tabular}
  \caption{Estimation of $\fbr$ comparing the NFW potential fitting to the caustic surface and model-independent caustic mass assuming $\fb = 0.5$ for all the 4 mass ranges. The left and right columns are without and with the assumption of the c-M relation. }
  \label{tab:fbr_est}
\end{table}
}

%%%%%%%%%%%%%%%%%%%%%%%%%%%%%%%%%%%%%%%%%%
\subsection{Consistency of mass profiles and the value of $\fb$}\label{sec:massprofiles}
%%%%%%%%%%%%%%%%%%%%%%%%%%%%%%%%%%%%%%%%%%
We now perform a more detailed comparison of the mass profiles obtained from the fitting to the caustic surfaces based on the NFW potential and the traditional caustic technique to estimate the value of  $\fb$. For this purpose, we estimated the caustic masses using $\fbr =0.5$, as reported in \cref{tab:GR_Masses} and, additionally, by constructing the caustic mass profiles assuming $\fbr\equiv \fb^{\rm NFW}$, using the posteriors obtained from the Bayesian analysis of the former NFW-based analysis employing \cref{eqn:fofr} and \cref{eqn:gbeta}. As shown in the top-left panel of \cref{fig:fbr2}, we find extremely good agreement between the mass profiles biases ($\Mnfwf/\Mcauf$) across all 4 mass ranges. In this approach, we compare and estimate the mass bias between the two techniques over the entire range of data available instead of limiting the comparison to $\Mnfw $ vs. $\Mcau$ alone. As one can notice, the bias is of the order of $\sim 10\%$, and completely in agreement with no bias at $\sim 1\sigma$ C.L. with the uncertainty level (estimated using $\Mnfwf$)\footnote{As we do here, it is appropriate to use the uncertainty either from the caustic technique or the NFW potential based fitting and not both at the same time, as the data utilized to estimate both uncertainties are the same. } shown by the shaded region. 

{Comparing the ratios of masses $M_{\mathrm{NFW}}/M_{\mathrm{Cau}(\fb =0.5)}$ at the respective $\rf$ of each mass range, we estimate the value of $\fb$ that is necessary to have no bias between the two mass estimates. As shown in the left column of \cref{tab:fbr_est}, we find that the mass ratios at $\rf$ are extremely consistent, while being $\sim 15\%$ larger than $\fb = 0.5$ at $\sim 1 \sigma$ C.L., in all mass ranges. A posterior importance sampling gives us a constraint on the value,} 
\begin{equation}
    \fb = 0.59\pm 0.05,
\end{equation}
which is an independent determination of the filling factor obtained from the real data, in contrast to the usual value inferred from $\LCDM$ based simulations. {However, these values are apparently in stark contrast to the recently estimated $\fb = 0.4\pm 0.1$ in \cite{Pizzardo:2023idp}, utilizing the illustrisTNG \citep{Nelson:2018uso} simulations based on $\LCDM$ cosmology. This could, in effect, provide us with a unique opportunity to also validate/falsify the $\LCDM$ model through comparisons with simulations. }

\begin{figure*}[!ht]

    \includegraphics[width=0.255\textwidth]{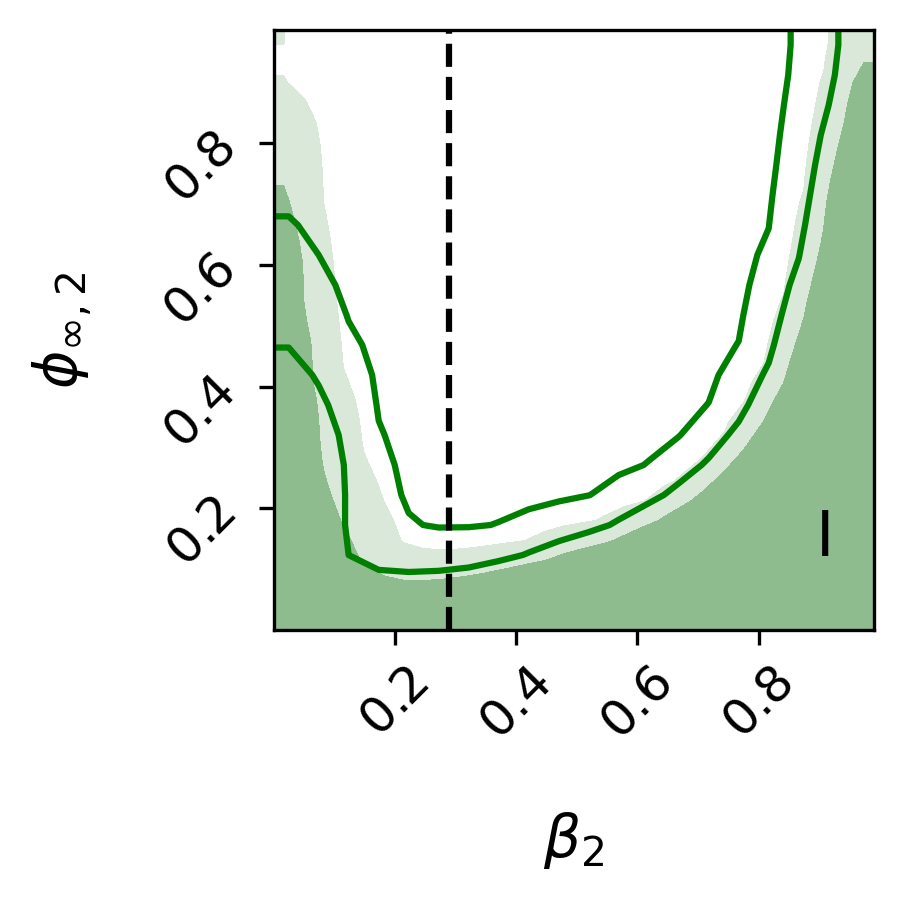}
    \hspace{-0.3cm}
    \includegraphics[width=0.255\textwidth]{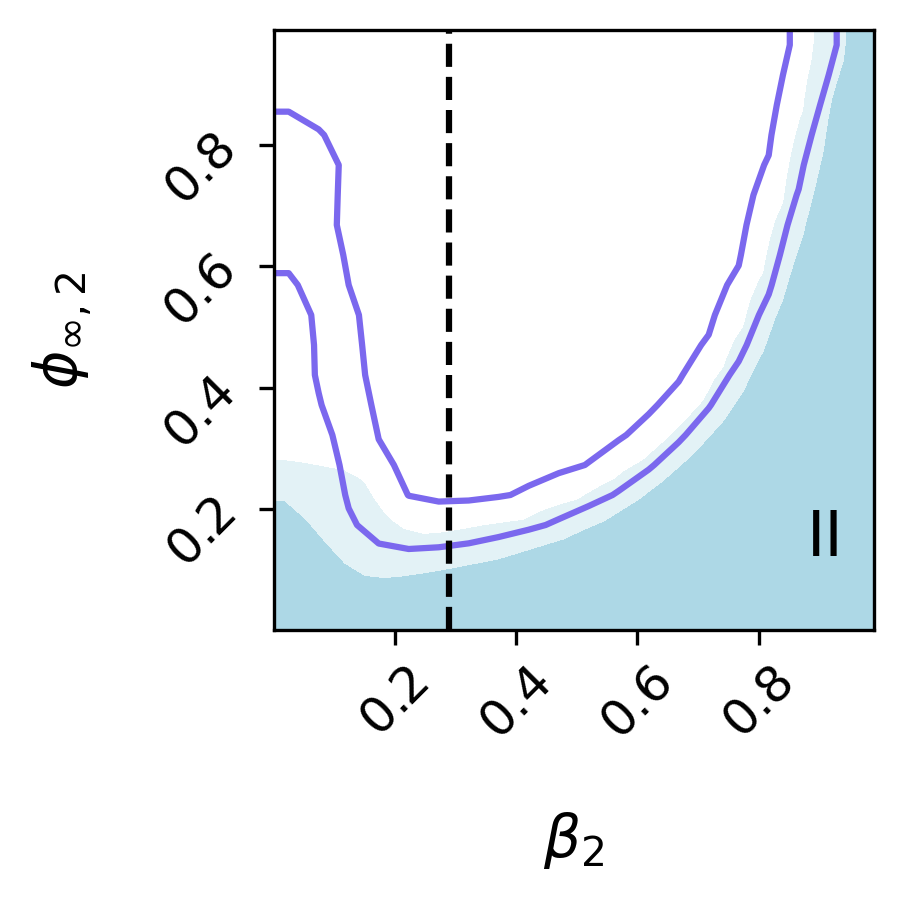}
    \hspace{-0.3cm}
    \includegraphics[width=0.255\textwidth]{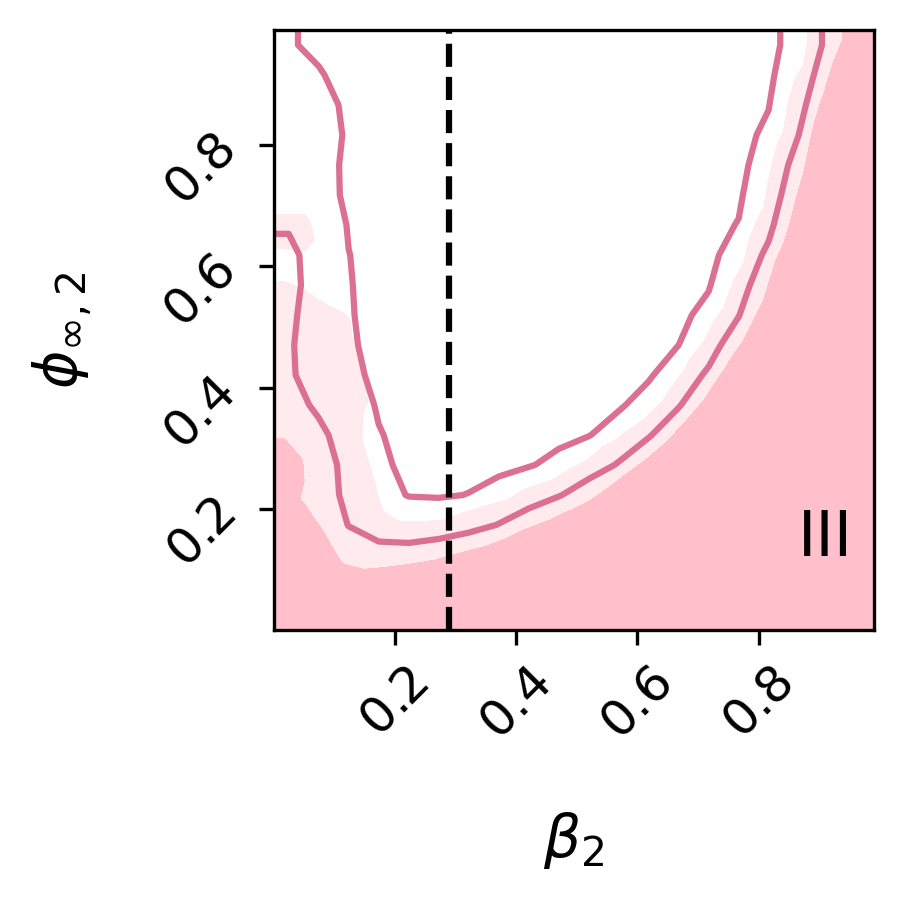}
    \hspace{-0.3cm}
    \includegraphics[width=0.255\textwidth]{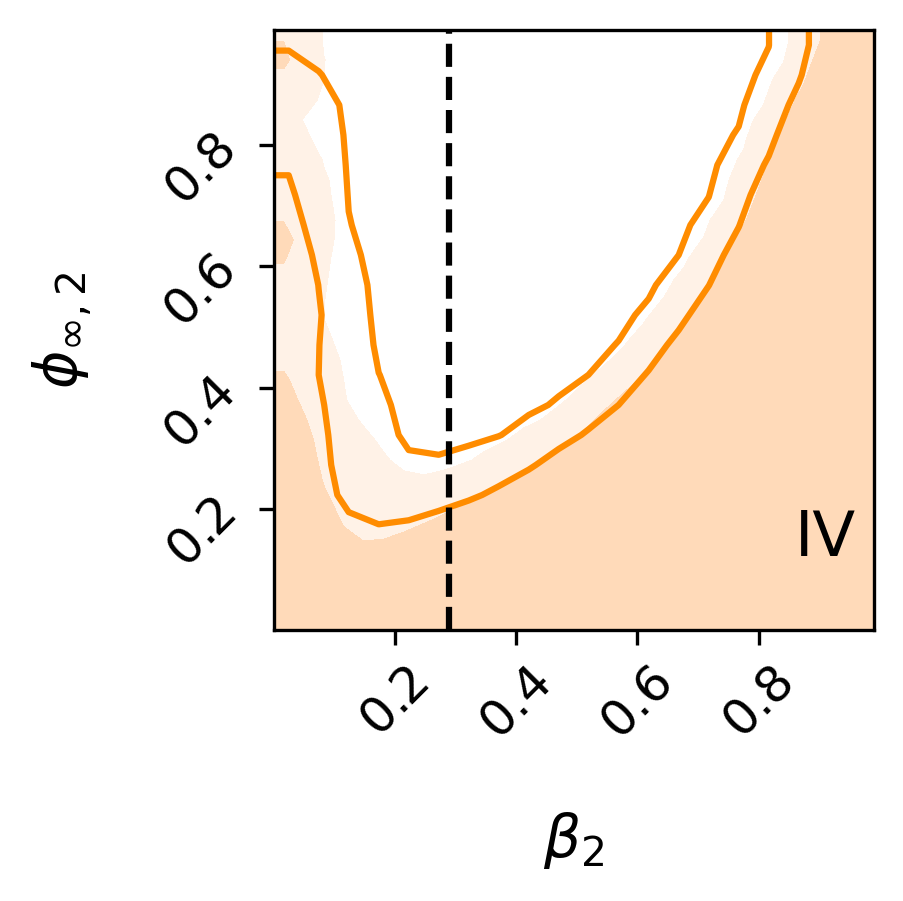}
  \caption{ We show the $95\%$ ($2\sigma$) and $99\%$ ($3\sigma$) C.L. posteriors for the modified gravity parameters now written in terms of, $\{\phitwo = 1- \exp{(\phi_{\infty} 10^{-4} M_{\rm Pl})},\, \betatwo = \beta/(1+\beta)\}$, in all 4 stacked caustics ranges of mass, going from range I to IV from left to right. We also show $\beta = \sqrt{1/6}$ as a vertical dashed line, corresponding to the scenario of $\fofR$ gravity. The unfilled contours represent the constraints obtained without the assumption of the c-M (M16) relation as a prior, while the filled contours include the c-M prior.}
  \label{fig:MG_params}
\end{figure*}

In a similar vein, to the analysis performed in \cite{Butt:2024jes}, we now compute $\fbr\equiv \fb^{\rm NFW}$ and estimate $\Mcauf(\fb^{\rm NFW})$, which we compare in the bottom-left panel. Once again, with excellent agreement between all 4 mass ranges, we note that the bias is of the order $1-\Mnfwf/\Mcauf(\fb^{\rm NFW}) \sim - 10\%$ level, essentially giving slightly higher caustic mass profiles. {Ideally, within this comparison one would expect the two mass profiles to agree with unity as the same underlying $\fbr$ is present in both techniques. In turn, the higher caustic mass profiles obtained here allow one to infer that the assumed constant anisotropy profile most probably provides the inferred bias. This, in effect, can be utilized to better constrain the anisotropy by making the two mass profiles match, say, at $\rf$, for each mass range. As already mentioned, we will explore this avenue in a future work where we shall assess the kinematics of the stacked clusters solving the Jeans equation.} We remain with the inference that the existing bias is a result of the assumption/fixing of the anisotropy profile in the NFW-potential based fitting and should be assessed accordingly. Overall, this comparison indicates a very good agreement between the two techniques, and no mass range dependent bias in the mass profiles.

On the other hand, in the right panel of \Cref{fig:fbr2}, we show a similar comparison when the c-M relation is imposed as a prior. Here, however, we find that the mass bias acquires a mild dependence on the mass range of the stacked clusters. For instance, we find a larger mass bias in the two lower mass ranges going up to $\sim 20\%$, while the two higher mass ranges show a bias of the order of $\sim 10\%$, indicating a decrease in the bias with increasing mass range. The bias is evident when comparing $\Mnfw/\Mcau(\fb = 0.5)$ at the respective $\rf$ of each mass range, {which clearly highlights the decreasing trend from $\fbr = 0.66\pm 0.09$ to $\fbr = 0.52\pm 0.06$, for the lowest and highest mass range, respectively}. We infer that the imposition of the c-M relation as a prior in the Bayesian analysis, while providing a better constraint on the concentration parameter, introduces a mass-dependent bias in the caustic mass profiles, especially within the radial range of $\rp < \rf$. Beyond the virial range, while the mass bias increases to $\sim 25\%$, it is completely consistent within the uncertainty level of the 4 mass ranges. This could, in effect, be a manifestation of the mass-anisotropy degeneracy, indicating that an appropriate assumption of the anisotropy should be different within different mass ranges. One could be well assessed by matching the ratios of the mass profiles to unity, as mentioned earlier. 

{\renewcommand{\arraystretch}{1.5}
\setlength{\tabcolsep}{8pt}

\begin{table*}[!ht] 
    \caption{Constraints on the mass and modified gravity parameters in the case of chameleon screening, shown at $68\%$ C.L. limits. The upper limits on $\phitwo (\fR)$ are shown at $95\%$ C.L. limit.  }
    \label{tab:CS_Masses}
    \centering
    \begin{tabular}{cccccccc}
        \hline 
        Range & c-M prior & \(\cnfw\) & \(\Mnfw\) &  $\phi_{\infty, 2}(\fR\,[10^{-5}])$ & \(\ln(\sigma_{\rm int})\)\\
        &  &  & \([\Munit]\)  &$\beta = \sqrt{\frac{1}{6}}$  & \\
        \hline
        \hline
        \multirow{2}{*}{I} & with & $3.79_{-0.81}^{+0.82}$ & $2.47_{-0.26}^{+0.21}$  & $<0.056(0.537)  $ &  $-3.26^{+0.58}_{-1.51}$ \\
        & w/o & $2.87^{+0.69}_{-0.60}$& $2.21^{+0.25}_{-0.26}$ & 
        $<0.057(0.541)  $&
         $-3.39^{+0.65}_{-1.64}$ & \\
        \multirow{2}{*}{II} & with & $4.31_{-0.62}^{+0.60}$ &$4.28_{-0.29}^{+0.26}$  &  $<0.084(0.807) $ &  $-2.43^{+0.20}_{-0.22}$ & \\
        &  w/o & $3.51^{+1.03}_{-0.81}$ &$3.93^{+0.44}_{-0.47}$ &  $<0.089(0.848)  $&
        $-2.39^{+0.19}_{-0.21}$&
        \\
        \multirow{2}{*}{III} &  with & $4.36_{-0.51}^{+0.49}$  & $5.79_{-0.33}^{+0.31}$  &  $<0.107(1.024)  $ & 
        $-2.26^{+0.18}_{-0.18}$&
        \\
        &  w/o & $4.45^{+1.24}_{-0.98}$ & $5.84^{+0.53}_{-0.56}$&  $<0.108(1.034)  $ &
        $-2.25^{+0.18}_{-0.18}$&\\  
        \multirow{2}{*}{IV} & with & $3.89_{-0.36}^{+0.38}$ & $9.61_{-0.56}^{+0.54}$  &  $<0.186(1.781)    $&
        $-1.71^{+0.15}_{-0.14}$ &
        \\
        &  w/o & $10.56^{+0.86}_{-0.93}$ & $5.32^{+1.49}_{-1.20}$& 
         $<0.175(1.675)   $&
        $-1.74_{-0.15}^{+0.15}$&\\

        \hline
    \end{tabular}
    \par
\end{table*}
}

%%%%%%%%%%%%%%%%%%%%%%%%%%%%%%%%%%%%%%%%%%
\subsection{Modified gravity}\label{sec:MGresults}
%%%%%%%%%%%%%%%%%%%%%%%%%%%%%%%%%%%%%%%%%%
We now focus on constraining the modified gravity parameters, $\{\phitwo, \betatwo\}$, in the case of chameleon screening using our 4 stacked caustic mass ranges. We show the posteriors for these modified gravity parameters obtained in the fully Bayesian analysis in \cref{fig:MG_params}. The MG parameters, along with the mass and concentration parameters, are reported in \cref{tab:CS_Masses}. We find that the constraints on the mass profile parameters are consistent with the standard GR case, both with and without the imposition of the c-M relation; in the former case, the differences are  even smaller. 

We find extremely good constraints on the exclusion posteriors of $\{\betatwo, \phitwo\}$, replicating the results found earlier in \cite{Terukina:2013eqa, Wilcox:2015kna, Boumechta:2023qhd} which use the hydrostatic datasets, along with mass priors from weak lensing in some cases. As can be already seen in \cref{fig:MG_params}, we find extremely good agreement in the constraints obtained with (filled) and without (unfilled) the prior cM. This, in turn, also validates the fact that the accuracy of determining the mass and concentration only mildly influences the constraints in the modified gravity model. A very interesting and important improvement we observe is that the degeneracy between the $\betatwo$ and $\Mf$ parameters that exists in the constraints obtained using the hydrostatics data as elaborated in  \cite{Boumechta:2023qhd} is no longer present when the potential is directly fitted using the caustic surface. This is clearly due to an advantage of the caustic formalism, where the assessments of modifications to gravity can be done directly on the potential. Whereas, in the context of hydrostatic equilibrium, the derivative of the potential undergoes an integration to obtain the thermal pressure, subsequently leading to the propagation of errors and yielding degeneracies in the final posteriors. Usually, the inclusion of a prior on the mass parameter within the hydrostatic analysis, for instance, using weak lensing observables, can aid in the elimination of said degeneracy \citep{Terukina:2013eqa, Boumechta:2023qhd}. In the current analysis using the caustic surface, we no longer have this need to assume an external prior from a complementary observable. This also validates what was found earlier in \cite{Butt:2024jes}, where utilizing the $\fbr$ estimated using hydrostatics as a prior in the caustic technique had reduced this degeneracy and improved constraints in the modified gravity parameter space. 

\begin{figure}
 \includegraphics[width=0.48\textwidth]{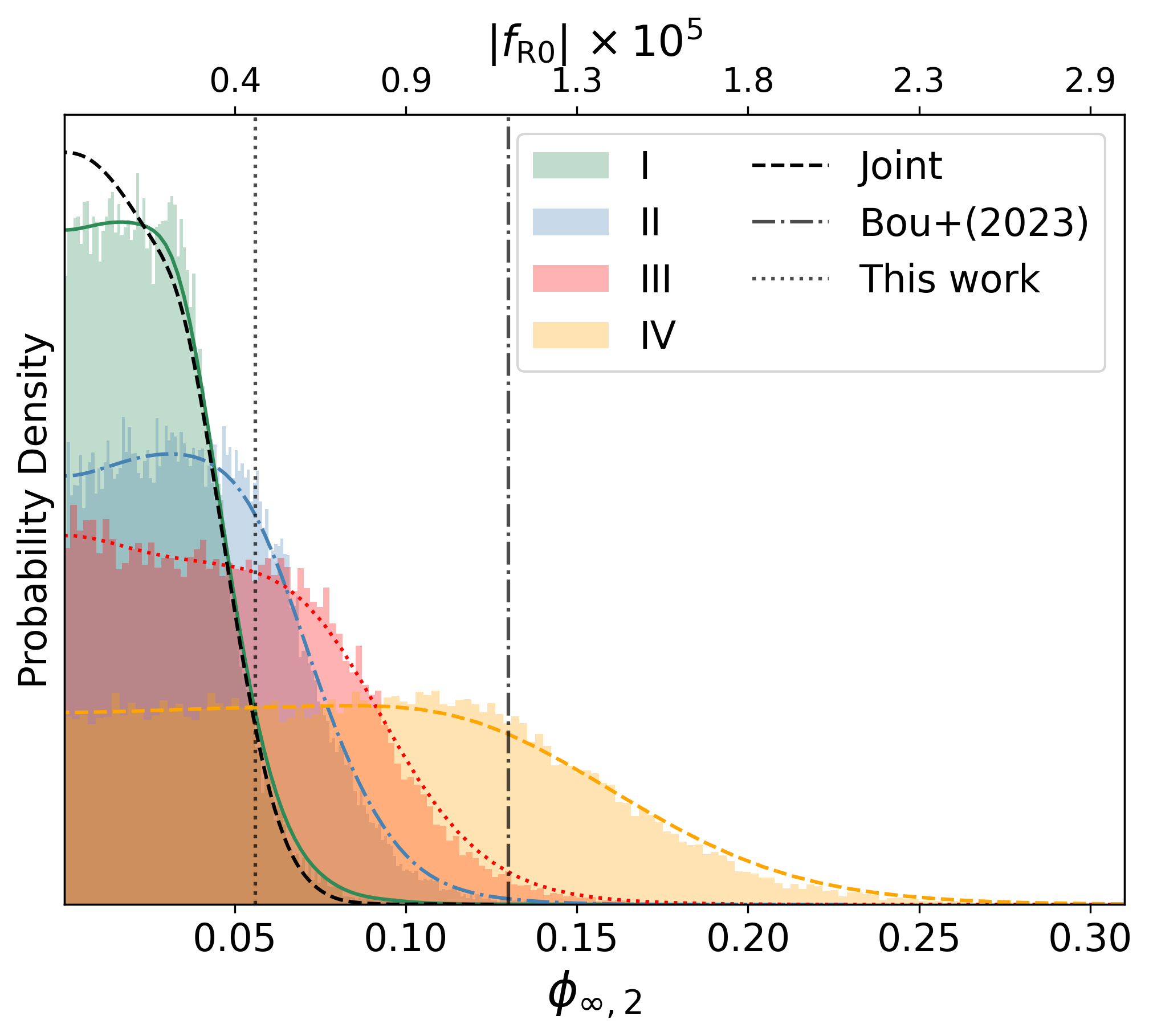} 
    \caption{Posteriors of $\phi_{\infty,2}$ for $\beta = \sqrt{1/6}$ fixed, which corresponds to the $\fofR$ gravity scenario. The vertical dash-dotted line represents the $95\%$ C.L. upper limit estimated in \cite{Boumechta:2023qhd} obtained using 5 stacked clusters along with the mass priors from weak lensing, whereas the dotted line represents the upper limit from our importance sampling analysis here (\cref{eqn:finalcons}). Here the posteriors are shown for the case with c-M assumption included. The black-dashed line shows the joint posterior importance sampled from the four mass bins.}
    \label{fig:prob_density_phi_1}
\end{figure}

In \cref{tab:CS_Masses}, we report the constraints obtained in the case of chameleon screening for the 4 mass ranges. In column 5 we show the $95\%$ C.L. upper limits on the asymptotic $\phitwo$, when $\beta = \sqrt{1/6}$, corresponding to the $\fofR$ scenario. As can be seen, the constraints become weaker as the mass of the stacked cluster increases. We find almost equivalent upper limits, both with and without the imposition of the c-M prior. Clearly, the constraints obtained in each of the first 3 mass ranges here are more stringent than those obtained earlier in the most recent analysis of the hydrostatic data of 5 X-COP clusters in \cite{Boumechta:2023qhd}. We show the posteriors of $\phitwo$ for $\beta = \sqrt{1/6}$ in \cref{fig:prob_density_phi_1}, and for comparison, the upper limit estimated in \cite{Boumechta:2023qhd} is shown as a dashed line. The tightest limit on $\phitwo$ found using the lowest mass stacked cluster is $\fR [ 10^{-6}] \leq 5.4 $, which is approximately twice as tight as the upper limit estimated using 9 stacked clusters $\fR [10^{-5}] \leq 1.2$, in \cite{Boumechta:2023qhd}. The advantage in our current analysis is two-fold i) the availability to stack caustic phase-spaces of several tens of clusters in a given mass range and ii) the lack of mass-coupling strength degeneracy in the current formalism. Finally, we perform a simple importance sampling of the posteriors obtained for the four stacked clusters {obtaining the joint constraint shown as the black dotted line in \cref{fig:prob_density_phi_1}. As can be seen, the lowest mass bin provides the most stringent constraint dominating over the joint likelihood making the constraint almost equivalent to the one obtained from the mass range-I alone. We report a final constraint of}
\begin{equation}\label{eqn:finalcons}
    \phitwo \leq 0.053 \implies \fR \leq 4.43 \times 10^{-6}, 
\end{equation}
shown as the dotted line in \cref{fig:prob_density_phi_1}. The results clearly demonstrate that the stacked caustic phase space is capable of constraining the gravitational potential of galaxy clusters, and hence the modifications from the standard scenarios.  

%%%%%%%%%%%%%%%%%%%%%%%%%%%%%%%%%%%%%%%%%%
\section{Conclusions} \label{sec:conclusions}
%%%%%%%%%%%%%%%%%%%%%%%%%%%%%%%%%%%%%%%%%%

The application of the caustic technique to study galaxy clusters beyond its standard application of solely estimating the mass could be of utmost importance. The caustic profile $\mathcal{A}(r)$ is one of the only available observable, aside from weak lensing, capable of assessing modifications to the gravitational potential at the outskirts of galaxy clusters. Earlier, we utilized the caustic technique in conjunction with hydrostatic data, in \cite{Butt:2024jes}, while in the current work, we utilize the former on its own. 

To this end, we have constructed a dataset of galaxy cluster caustic surfaces by stacking the phase spaces of several well-observed individual clusters in the redshift range of $z\in\{0.01, 0.1 \}$. The advantage of stacked objects is that they can be utilized to study gravity on cluster scales, well beyond the virial radius. We validate our stacking method by comparing the masses obtained using the model-independent (calibration-dependent) caustic technique \citep{Diaferio_1997} and our model-dependent (NFW mass profile) mass constraints. By contrasting the NFW assumption based mass profiles and the model-independent mass profiles (assuming $\fb = 0.5$), we estimate the value of the filling factor to be $\fb = 0.59\pm 0.05$. This estimate is solely a data-based constraint on the filling factor, without resorting to any simulations. This in turn provides a unique avenue to test $\LCDM$, by comparing the filling factor values obtained from simulations and data directly.

As a validation and necessary sanity check, we performed the analysis as implemented for the stacked clusters, on the cluster, A2029 (see App. \ref{sec:A2029}). We find very good consistency between the inferences made using the stacked clusters and the cluster A2029. We also show that the analysis with a fixed anisotropy profile is completely consistent with the constraints obtained when marginalizing over the anisotropy profile parameters. In fact, the anisotropy profile parameters are strongly degenerate with the concentration parameter rather than the mass itself, where the constraints on the former are vastly improved when the c-M relation is imposed as a prior. In addition, the constraints on the mass are only mildly enlarged when marginalizing over the anisotropy parameters. 

We then proceed to utilize the dataset to constrain modified gravity, specifically the chameleon screening model \citep{Khoury:2001bz} as a test case here, finding vastly improved constraints on the field parameters on the clusters' scale, with respect to those present in the literature \citep{Terukina:2013eqa, Wilcox:2015kna, Boumechta:2023qhd, Pizzuti:2020tdl, Pizzuti:2022ynt}. Performing an importance sampling analysis of the constraints on the $\fofR$ model, using each of the 4 stacked caustic surfaces, we estimate a value of $\fR \leq 4.43 \times 10^{-6}$. This is the tightest constraint obtained yet, using galaxy clusters, albeit obtained while fixing the kinematics of the galaxy cluster. In addition, it is a caustics-only assessment of any modified gravity scenario, for the first time. Although we have remained with the standard NFW assumption for the dark matter density profile, an extended analysis to assess different mass models \citep[see for instance][]{Pizzuti:2024vjz} could be an interesting avenue. 

Needless to say, also within the GR formalism, assessing the nature of dark matter in the outer regions of galaxy clusters can be facilitated by the datasets constructed here. And can be directly utilized to constrain other modifications to gravity and varied dark matter scenarios \citep{Benetti:2023vxy, Gandolfi:2023hwx, Cardone2020Aug, Laudato:2021mnm, Pizzuti:2024stz, Zamani:2024oep, Zamani:2024qbx, Salzano:2016udu, Benetti:2024nxr, Hodson:2017dxw, Famaey:2024uow, Bouche:2024qhy}, especially at the outskirts of galaxy clusters, which is a limitation of other observables (besides weak lensing) such as the SZ-pressure and X-ray temperature \citep{Ettori:2018tus}, which only extend, at most, slightly beyond the virial radius. Also, the caustic surface as an observable, is independent of the dynamical nature of galaxy clusters, which allows for the construction of highly constraining datasets. In a {future} extension of the current analysis, we also intend to study the kinematical behavior of the stacked clusters to constrain the anisotropy profile, in order to generalize the formalism beyond the current analysis.

%%%%%%%%%%%%%%%%%%%%%%%%%%%%%%%%%%%%%%%%%%
\section*{Acknowledgments}   
%%%%%%%%%%%%%%%%%%%%%%%%%%%%%%%%%%%%%%%%%%

BSH is supported by the INFN INDARK grant and acknowledges support from the COSMOS project of the Italian Space Agency (cosmosnet.it). CB acknowledges support from the COSMOS project of the Italian Space Agency (cosmosnet.it), and, together with AD, the INDARK Initiative of the INFN (web.infn.it/CSN4/IS/Linea5/InDark). AL has been supported by: European Union - NextGenerationEU under the PRIN MUR 2022 project n. 20224JR28W "Charting unexplored avenues in Dark Matter"; INAF Large Grant 2022 funding scheme with the project "MeerKAT and LOFAR Team up: a Unique Radio Window on Galaxy/AGN co-Evolution"; INAF GO-GTO Normal 2023 funding scheme with the project "Serendipitous H-ATLAS-fields Observations of Radio Extragalactic Sources (SHORES)"; project ``Data Science methods for MultiMessenger Astrophysics \& Multi-Survey Cosmology'' funded by the Italian Ministry of University and Research, Programmazione triennale 2021/2023 (DM n.2503 dd. 9 December 2019), Programma Congiunto Scuole; Italian Research Center on High Performance Computing Big Data and Quantum Computing (ICSC), project funded by European Union - NextGenerationEU - and National Recovery and Resilience Plan (NRRP) - Mission 4 Component 2 within the activities of Spoke 3 (Astrophysics and Cosmos Observations).

\bibliographystyle{aasjournal}
\bibliography{bibliography}{}

\begin{appendix}
Here we present the discussion pertaining to a few systematic differences in the final constraints that could arise due to our choice of the fitting range and the assumptions on the anisotropy profile. We also present an analysis of the cluster A2029, which acts as a validation of the analysis performed using the stacked clusters. 

%%%%%%%%%%%%%%%%%%%%%%%%%%%%%%%%%%%%%%%%%%
\section{Systematics}\label{app:systematics} 
%%%%%%%%%%%%%%%%%%%%%%%%%%%%%%%%%%%%%%%%%%   

As we have mentioned in the main text, we choose the radial range of $0 \leq \rp/\Mpc \leq 4.0$ as our primary range of the caustic surface as the dataset to perform the Bayesian analysis, while scaling the error on the data points $\rp<0.5$, to have twice as large error. This under-weighting of the caustic data in the inner regions of the cluster is done to avoid any biases in our fitting of the NFW potential to the caustic surface, where the former tends to rise steeply in the inner regions of the cluster and the latter usually decreases or remains flat. This effect is particularly more evident in the higher mass stacked clusters as can be seen in the bottom panels of \cref{fig:caustic_surface}. 

\begin{figure}
    \centering
    \includegraphics[width=0.497\linewidth]{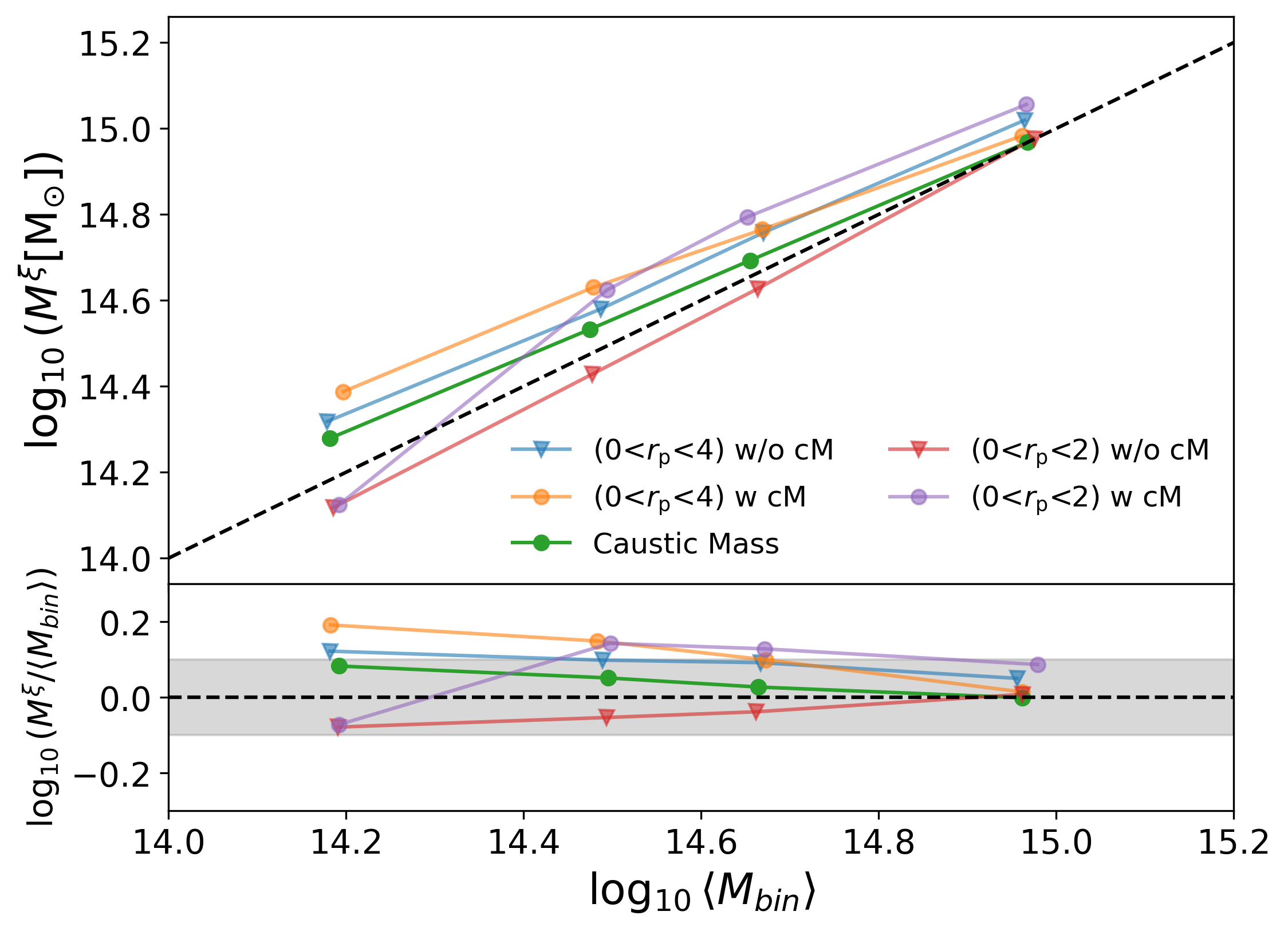}
    \includegraphics[width=0.497\linewidth]{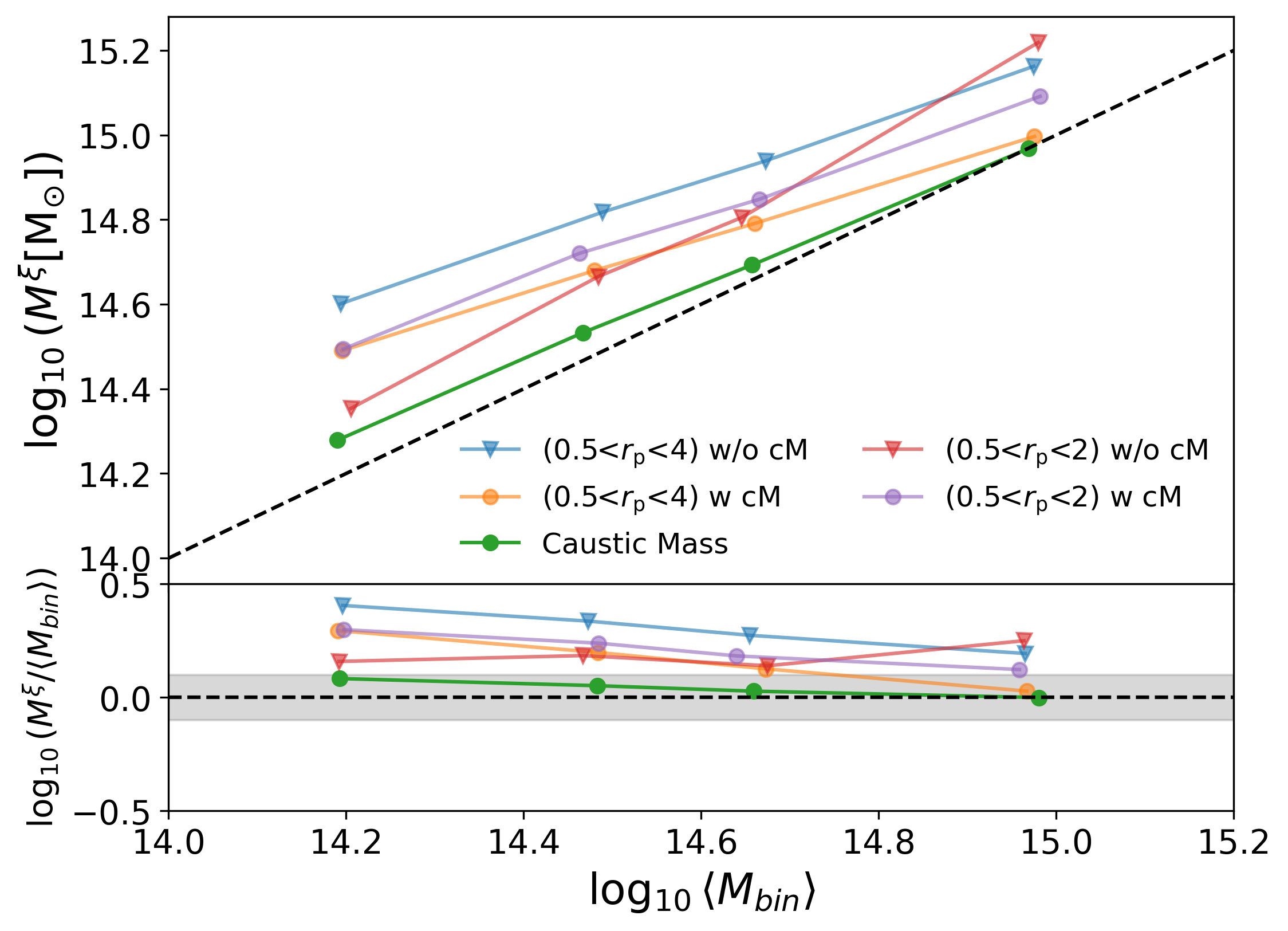}
    \caption{\textit{Left}: We show the comparison of the mass bias when utilizing different radial ranges to fit the NFW potential. \textit{Right}: Same as the left panel but with the lower limit of $\rp>0.5$ imposed on the data.}
    \label{fig:mass_bias}
\end{figure}

However, we also test for variations when using the upper limit of $\rp \leq 2\, \Mpc$. We find systematically that the masses are overestimated when the radial range is limited to the latter. While the reduced $\chi^2$ remains equivalent for both ranges, this difference is clearly driven by the steeper slope of the caustic profile in the inner regions of the cluster alone when the radial range is limited to $\rp \sim 2 \, \Mpc$, with no compensation provided by the full range. As can be seen in the left panel of \cref{fig:mass_bias}, we find that the caustic mass estimated in each of the mass ranges is in very good agreement with the expected mass bias, as presented earlier in \cite{Gifford:2016plw}, always within the $10\%$ limit. In the left panel of \cref{fig:mass_bias}, we compare the NFW masses estimated while taking into account the entire range of the caustic surface and the NFW masses estimated taking into account the radial range of $0 \leq \rp/\Mpc \leq 2.0$. We find that the mass bias is driven by the inner regions of the cluster, which is not compensated by the outer regions. This is clearly seen in the right panel of \cref{fig:mass_bias}, where we impose the lower limit of $\rp>0.5\, \Mpc$ on the data.

%%%%%%%%%%%%%%%%%%%%%%%%%%%%%%%%%%%%%%%%%%
\subsection{c-M relation}\label{app:cM}
%%%%%%%%%%%%%%%%%%%%%%%%%%%%%%%%%%%%%%%%%%
We assess the effect of including the c-M relation in our method, as detailed in \Cref{sec:cM} and further described in \Cref{sec:NFW Fitting}. Firstly, we find that the c-M relation has only a mild effect on the NFW mean mass estimates of each of the stacked clusters, as shown in \cref{fig:GR_contours}. However, it is interesting that collectively the constraints from the 4 mass ranges without the imposition of the c-M prior (see left panel of \cref{fig:GR_contours}) follow a trend that is in contrast to the well-established c-M relations \citep{Merten_2015, Dutton:2014xda, Child_2018}. For instance, fitting the M16 relation to the posteriors of the 4 stacked masses in our analysis w/o the c-M relation, we find the parameters $\{A, B, C\} = \{11.27, -2.58, -0.44\}$, which are to be contrasted against the values $\{3.66, -0.14, -0.32\}$ that are presented in \cite{Merten_2015}. When we repeat the same exercise with the posteriors that are obtained with the imposition of the c-M relation (right panel of \cref{fig:GR_contours}), the mean c-M relation is given by $\{A, B, C\} = \{1.32, 3.56, -0.02\}$, which is within $\sim 1\sigma$ dispersion of the M16 relation. This set of parameter values also improves the agreement between the M16 and D14 \citep{Dutton:2014xda} relations, making the former flatter and hence, similar to the latter. As elaborated in the main text, we shall explore this further alongside the marginalization on the anisotropy profile of the stacked clusters.

%%%%%%%%%%%%%%%%%%%%%%%%%%%%%%%%%%%%%%%%%%
\subsection{Parametrization of $g_{\beta}$}\label{app:beta}
\begin{figure}[h]
\centering
    \includegraphics[width=0.5\linewidth]{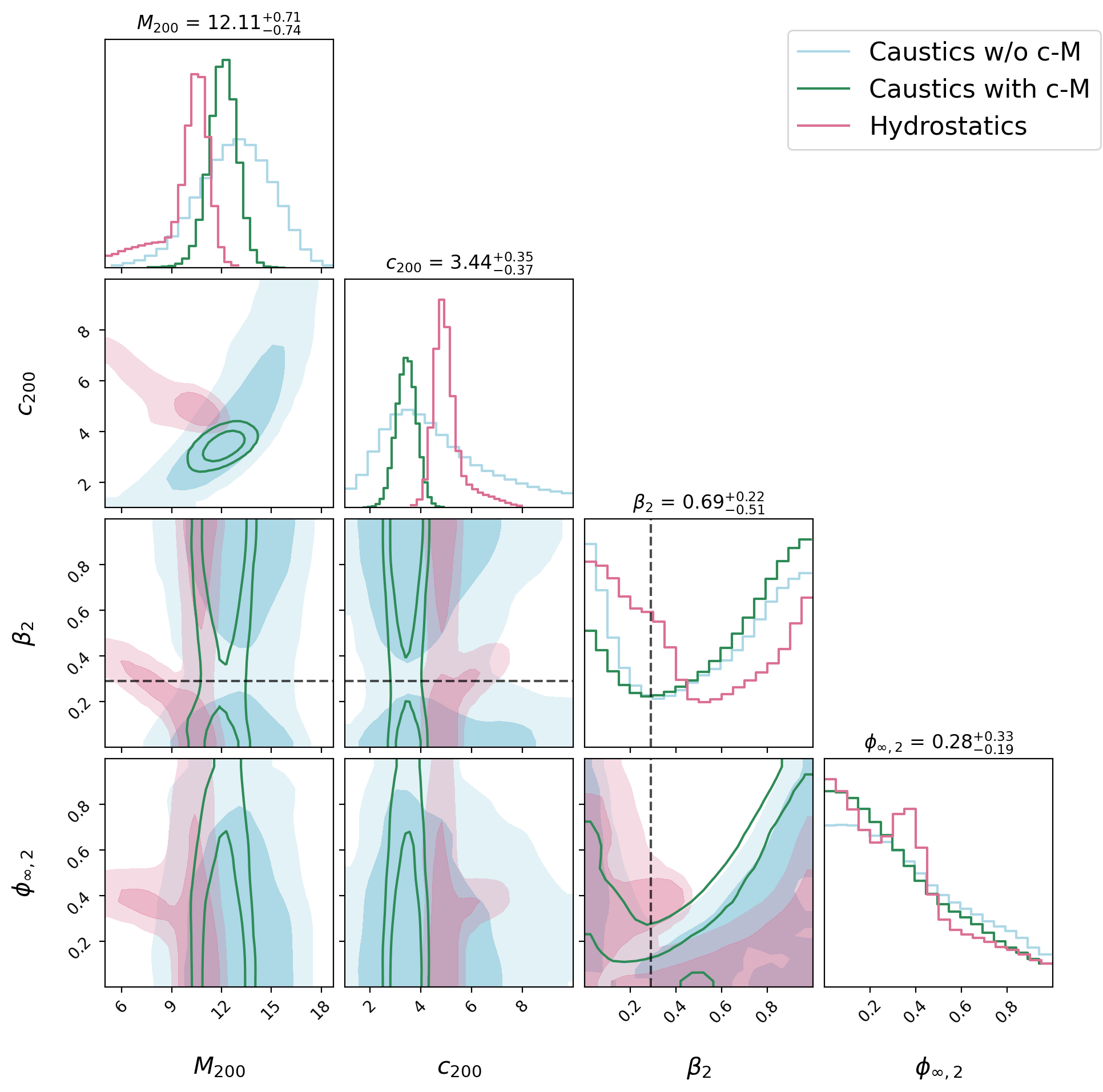}
    \hspace{-0.2cm}
    \includegraphics[width=0.5\linewidth]{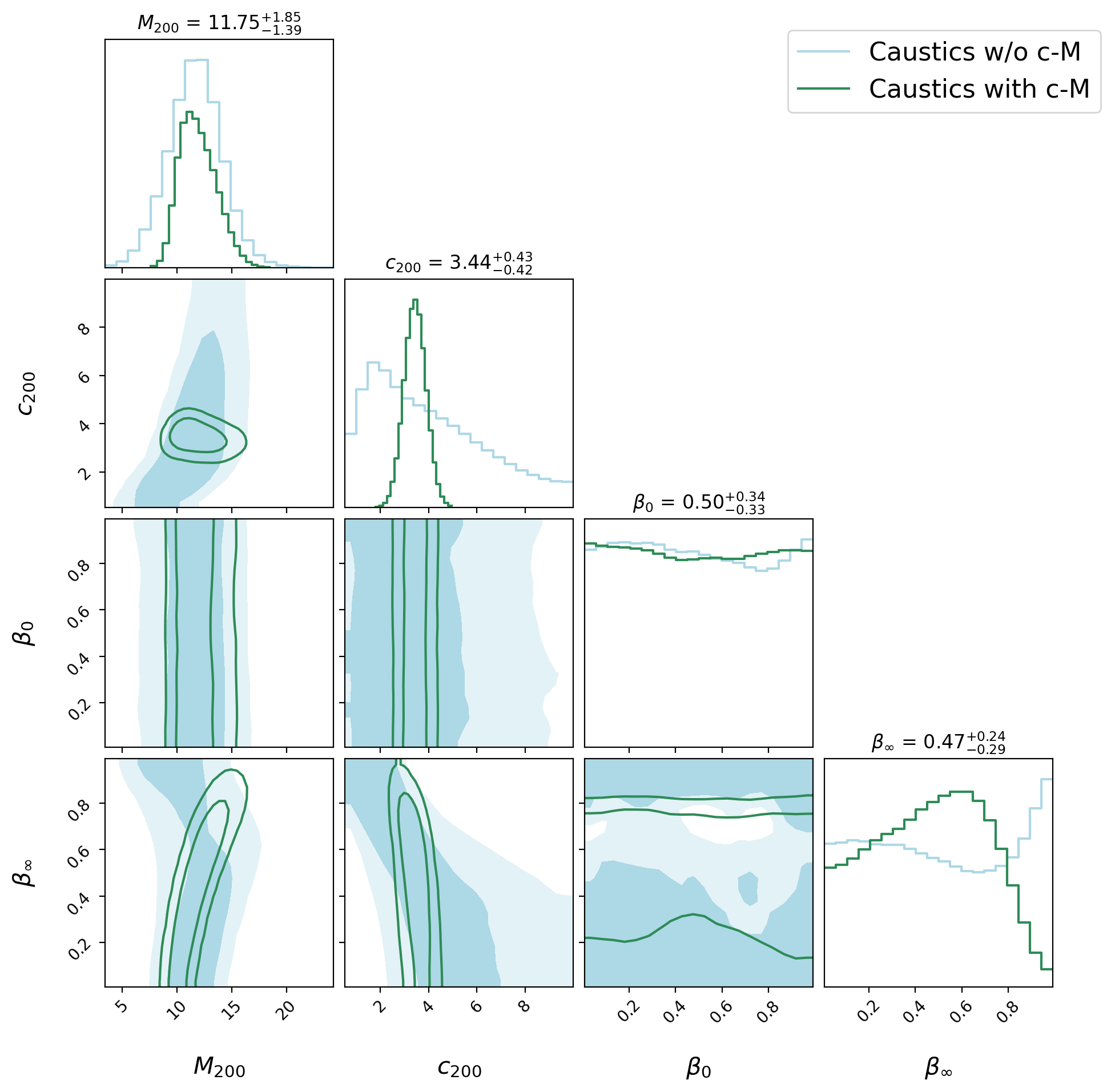}
  \caption{\textit{Left}: We show the posteriors for the $68\%$ and $95\%$ C.L. for the mass, concentration, and the chameleon 
  screening parameters, $\phi_{\infty,2}$ and $\beta_{2}$, included. Here, the anisotropy profile is kept fixed as in the main analysis. Note that, in the main text, the constraints on the modified gravity parameters (\cref{fig:MG_params}) are present for the $95\%,\, 99\%$ C.L. limits. \textit{Right}: We show the posteriors for the $68\%$ and $95\%$ C.L. for the mass, concentration, $\beta_{0}$ and $\beta_{\infty}$ parameters in the standard GR case with the NFW mass profile and assuming the generalized Tiret's anisotropy profile \cref{eqn:gbeta},  with and without the c-M prior for A2029. }
  \label{fig:A2029}
\end{figure}

%%%%%%%%%%%%%%%%%%%%%%%%%%%%%%%%%%%%%%%%%%
As mentioned in \Cref{sec:NFW Fitting}, we assume $\beta(r)$ to be the Tiret's model \citep{Tiret2007}, where $\beta_{\infty}$ is a parameter of the model. As in our previous analysis using the caustic method in \cite{Butt:2024jes}, we fix $\beta_{\infty}$ to be 0.5. In the current analysis, we compare the results obtained by fixing $\beta_{\infty}$ to those obtained by taking it to be a free parameter in our model and marginalizing over the allowed range $0 < \beta < 1$.\footnote{We ignore the possibility that the value of $\beta(r)$ can 
also be negative and extend to $\beta (r)\to-\infty$, if the mean radial velocity of the galaxies $\langle v_{r}^2\rangle$ is zero in a shell volume centered on $r$. This is however non-physical and sets the lower limit of $\gbr \to 2 $ for $\beta \to -\infty$. In the case of A2029 it has been shown that the anisotropy profile tends to be negative ($\beta(r) \to -0.4 $) in the inner regions ($r \lesssim r_{500}$) of the cluster \cite{Li:2023zua}.} Note that the results obtained in both cases are in very good agreement,  showing a slightly larger error in the mass estimates and appropriate correlation with the concentration parameter. Here we demonstrate this in the case of A2029 (\cref{fig:A2029}), as elaborated in the next section. 

%%%%%%%%%%%%%%%%%%%%%%%%%%%%%%%%%%%%%%%%%%
\section{A2029 Validation}\label{sec:A2029}
%%%%%%%%%%%%%%%%%%%%%%%%%%%%%%%%%%%%%%%%%%
In this section, we present a validation of our formalism, utilizing the cluster A2029. Previously, we used this cluster in \cite{Butt:2024jes} (see also \cite{Butt:2024civ}), to study the joint analysis of hydrostatic data \citep{Ettori:2018tus} and caustic data \citep{Sohn_2017}. Briefly, in \cite{Butt:2024jes}, we have utilized the Monte Carlo samples from the Bayesian analysis of hydrostatic data performed in \cite{Boumechta:2023qhd}, to estimate the filling factor $\fbr$ instead of fixing it to a constant value. This allowed us to accurately estimate the mass of the cluster using the caustic technique without relying on the assumption $\fbr={\mathrm const}$, consequently reducing the mass bias between the two techniques. In turn, we performed an importance sampling-based joint analysis, obtaining constraints for the mass, and the chameleon screening parameters. Therein, we have shown that this allows us to improve the constraints on the mass estimation and additionally on modified gravity parameters. In the current work, following the methodology outlined in the main text, we only use the caustics' information of A2029 to estimate the mass and the modified gravity parameters.

In the right panel of \cref{fig:A2029}, we show the posteriors in the analysis of the standard GR scenario marginalizing over the $\beta_{0}$ and $\beta_{\infty}$ parameters, assuming the generalized Tiret's anisotropy profile (\cref{eq:betaT}) for estimating $\gbr$  (\cref{eqn:gbeta}). We find that the mass estimates are in very good agreement with the caustic mass, both with and without the inclusion of the c-M relation as a prior. The constraints on the $\beta_{0}$ and $\beta_{\infty}$ parameters are in good agreement with the constraints obtained in the main analysis where they have been assumed to be constants, $\beta_0 = 0.0$ and $\beta_{\infty} = 0.5$. This, in effect, validates our approach in the main analysis. In this context, we also find that the inclusion of the c-M relation as a prior has a mild effect on the mass estimates, as shown in the right panel of \cref{fig:A2029}, while being able to improve limits on the anisotropy parameters: for instance $\beta_{\infty} \lesssim 0.8$ at $2\sigma$ C.L. and a median of $\beta_{\infty} \sim 0.6$ and mean of $\beta_{\infty} \sim 0.5$; in fact, we assume this value and keep it constant in our main analysis with the stacked clusters. We obtain similar conclusions in the case of the stacked clusters, where no preference/convergence for the $\{\beta_0, \beta_{\infty}\}$ parameters was found. As already mentioned in the main text, in future analyses we intend to assess the anisotropy profile, solving the Jeans equation \citep{Gifford_2013Ani, Falco:2013zya, Stark:2017vfa,  Pizzuti:2020tdl} and therefore, improving the overall formalism.

In the left panel of \cref{fig:A2029}, we show the posteriors in the case of chameleon screening, assuming the NFW mass profile for A2029. We find that the mass estimates are in very good agreement with the caustic mass, both with and without the inclusion of the c-M relation as a prior. It is interesting to note that there is a negative correlation between the mass and the concentration parameter within the hydrostatic analysis, which is aligned along the slope of the c-M relation (e.g., M16). This is in contrast to the constraints from the caustic analysis, which shows a positively-correlated degeneracy, and is a valid observation for both cases with and without the c-M relation as a prior, where the impact of the c-M relation is clearly evident. We infer that the degeneracy present in the hydrostatic analysis is due to the propagation of the uncertainties when integrating the derivative of the potential, when fitting against the SZ-pressure ($\Psz$) or the X-ray temperature ($\Tx$) data. Conversely, in our current formalism we directly fit the potential to the caustic surface, minimizing the propagation of the uncertainties, where the anisotropy information in $\gbr$ only acts as a normalization. This makes a strong case for the importance of utilizing the caustic technique in conjunction with the hydrostatic data to estimate the masses of galaxy clusters and the modified gravity parameters. Note that the concentration parameter $c_{200}$, as estimated in the hydrostatic analysis, is larger in comparison to that estimated in the caustic analysis assuming the c-M relation. 

\end{appendix}
\end{document}